\documentstyle[seceq,supplement,epsf,wrapft]{ptptex}
\pubinfo{No. 133, 1999}  
\notypesetlogo  
\preprintnumber[3cm]{
YITP98-20 \\ March 1998}

\title{
\Large \bf Distances in Inhomogeneous Cosmological Models}

\author{%
Kenji {\sc Tomita}$^{1,}$\footnote{ E-mail address: tomita@yukawa.kyoto-u.ac.jp},
Hideki {\sc Asada}$^{2,}$\footnote{ E-mail address: asada@phys.hirosaki-u.ac.jp}
and Takashi {\sc Hamana}$^{3,}$\footnote{ E-mail address: hamana@astr.tohoku.ac.jp}    
}

\inst{%
$^1$ Yukawa Institute for Theoretical Physics,
Kyoto University, Kyoto 606-8502, Japan\\
$^2$ Faculty of Science and Technology, Hirosaki University, 
Hirosaki 036-8561, Japan\\
$^3$ Astronomical Institute, Tohoku University, Sendai 980-8578, Japan
}

\recdate{%
February 10, 1999
}

\abst{%
Distances play important roles in cosmological observations,
especially in gravitational lens systems, but there is a problem in
determining distances because they are defined in terms of
light propagation, which is influenced gravitationally by the
inhomogeneities in the universe. In this paper we first give the 
basic optical relations and the definitions of different distances in 
inhomogeneous universes. Next we show how the observational relations depend
quantitatively on the distances. Finally, we give results for the 
frequency distribution of different distances and the shear effect 
on distances obtained using various methods of numerical simulation.
}

\begin{document}

\maketitle

\section{Introduction}

In optical relations among observed quantities, distances such as the
luminosity distance and the angular diameter distances play an
important role. They are clearly defined in the homogeneous
Friedmann-Lemaitre-Robertson-Walker model (Weinberg,\cite{rf:weintx} 
Schneider et al.\cite{rf:sef}) owing to the simple nature of light 
propagation in this case. In inhomogeneous
universes, however, their behavior is complicated, due to gravitational
lens effect which implies that light rays are deflected gravitationally
by an inhomogeneous matter distribution. On the other hand, we also use
 distances to interpret the structure of gravitationally lensed
systems.

To correctly treat distances in inhomogeneous universes, it is
necessary first to have a reasonable formulation for the dynamics
describing local matter motion and 
optics and clarify the validity condition of the formulation. A set of 
fluid dynamical equations and the Poisson equation in the cosmological 
Newtonian approximation was introduced and discussed by 
Nariai\cite{rf:nari60} and Irvine\cite{rf:irv65} under the conditions
\begin{equation}
  \label{eq:int1}
 \vert \Phi \vert \ll 1, \ (v/c)^2 \ll 1, \ L/L_H \ll 1,
\end{equation}
where $\Phi, \ v, L$ and $L_H$ are the Newtonian gravitational potential,
matter velocity, the characteristic size of inhomogeneities and the
horizon size $\approx ct$,
respectively, and the spacetime is expressed as 
\begin{equation}
  \label{eq:int2}
 ds^2 = -(1 +2\Phi)c^2 dt^2 +(1 -2\Phi) a^2(t)[d\chi^2 +\sigma^2(\chi) 
 d\Omega^2],
\end{equation}
where $a(t)$ is the scale-factor, $\sigma (\chi) = \sin \chi, \chi,
\sinh \chi$ for the background curvature $k = 1, 0, -1$, respectively, 
and $d\Omega^2 = d\theta^2 + \sin^2 \theta \varphi^2$.  
The above fluid dynamical equations can describe the nonlinear local
motion, while the gravitational field is linear with respect to
$\Phi$. The extension of the above cosmological Newtonian treatment to 
a post-Newtonian treatment was performed by Futamase,\cite{rf:futa88}
Tomita\cite{rf:tpost88} and  Shibata and Asada.\cite{rf:shib96} 
Futamase showed that the condition 
\begin{equation}
  \label{eq:int3}
 \epsilon^2 /\kappa \ll 1 \quad ( \epsilon^2 \sim \Phi \ {\rm and } \ \kappa
 \sim L/L_H)
\end{equation}
is necessary for the higher-order expansion to be possible and
formulated the spatial averaging and the back-reaction to the background.
Moreover, Futamase and Sasaki\cite{rf:fs89} investigated the validity
of light propagation in the cosmological Newtonian iterative 
approximation and discussed the distance problem.

In an empty region as the limiting inhomogeneous case, distances
exhibit behavior very different from those in homogeneous models
(the Friedmann distances). In the special case without tidal
shear from surrounding regions, the so-called Dyer-Roeder angular
diameter distance was derived by Zel'dovich,\cite{rf:zel64} Dashevskii 
and Slysh\cite{rf:dash66}
and Dyer and Roeder.\cite{rf:dyroe72}\tocite{rf:dyroe73}
In a non-empty region with a constant matter density $\rho_m$ but no 
tidal shear, we have the generalized
Dyer-Roeder distance with the clumpiness (or smoothing) parameter
$\alpha$, which is defined as 
\begin{equation}
  \label{eq:int4}
\rho_m/\rho_{\rm F} = \alpha = {\rm const}
\end{equation}
for the Friedmann density $\rho_{\rm F}$. The observational results derived
from the optical relations depend on whether we use
the Friedmann distances or the Dyer-Roeder distance. Quantitative
estimates for these difference and the effect of the cosmological
constant  have been studied by Fukugita et al.\cite{rf:fuk92}
and Asada.\cite{rf:asad98} On the other hand, it is important to determine
 what distances are most applicable and what value of the above 
parameter $\alpha$ is best, in realistic inhomogeneous models.
Kasai et al.\cite{rf:kasa90} and Watanabe and Tomita\cite{rf:wt90}
numerically calculated the frequency distribution for generalized
distances in simple models in which particles are distributed randomly.
Recently, Tomita\cite{rf:tom98c} derived this distribution in more realistic 
inhomogeneous models generated using the $N$-body simulation with the 
CDM spectrum. The general result is that the average value of $\alpha$ is 
nearly 1 and its dispersion decreases with the increase of the
redshift $z$, though it is $\sim 1$ for $z = 0.5$.

Another interesting topic is that involving the role of the shear 
term and the Ricci and Weyl focusing terms (in the optical scalar 
equation\cite{rf:sachs61})
 in the behavior of distances. To this time, the shear
effect has been discussed by Weinberg,\cite{rf:wein76} Watanabe et
al.,\cite{rf:wts92} Watanabe and Sasaki,\cite{rf:ws90} and
Nakamura,\cite{rf:naka97} and its focusing effect has recently been 
studied numerically by Hamana using a Monte Carlo simulation, taking 
into account small-scale inhomogeneities.

In this review paper, basic optical relations and the definition of
distances are first given in \S 2, the lensing relations
in the generalized Dyer-Roeder distance are derived in \S 3 (by Asada),
the statistical behavior of distances analyzed in  numerical
simulation is described in \S 4 (by Tomita), and the shear and focusing
effects are discussed in \S 5 (by Hamana). 

\newpage
\section{Optical scalars and the definition of distances}

\subsection{Geometry of ray bundles}
Rays are expressed as $x^\mu =  x^\mu(y^i, v)$, where $v$ is an
affine parameter. For each ray, the $y^i$ have constant values
($C^i$). The wave vector $k^\mu = \partial x^\mu/\partial v$ satisfies 
the null  condition and null-geodesic equation
\begin{equation}
  \label{eq:opt1}
k^\alpha k_\alpha = 0, \quad {k^\alpha}_{;\beta} k^\beta = 0.
\end{equation}
For two rays with $y^i = C^i$ and $y^i = C^i + \delta C^i$, the
connection vector is 
\begin{equation}
  \label{eq:opt2}
\delta x^\mu = ({\partial x^\mu/\partial y^i}) \delta C^i.
\end{equation}
If we define a dot differentiation by  
$(m^\alpha)^\cdot = {m^\alpha}_{;\beta} k^\beta$ \
%
 for an arbitrary vector $m^\mu$, we obtain from Eq. (\ref{eq:opt2})
\begin{equation}
  \label{eq:opt4}
(\delta x^\alpha)^\cdot = {k^\alpha}_{;\beta} \delta x^\beta,
\end{equation}
and then (Jordan et al.,\cite{rf:jord}  Sachs\cite{rf:sachs61})
\begin{equation}
  \label{eq:opt5}
(k_\alpha \delta x^\alpha)^\cdot = 0.
\end{equation}

Now let us consider the situation in which on a screen (at a point
P$_{\rm o}$) an observer sees the shadow formed  by a source object 
(at a point 
P$_{\rm s}$). The connection vector $\delta_\bot x^\alpha$ vertical to
$u^\alpha$ and $k^\alpha$ at the point P$_{\rm s}$ is  
\begin{equation}
  \label{eq:opt6}
\delta_\bot x^\alpha = h^\alpha_\beta \delta x^\beta,
\end{equation}
where
\begin{equation}
  \label{eq:opt7}
h^\alpha_\beta = \delta^\alpha_\beta - {k^\alpha k_\beta \over
(k^\gamma k_\gamma)^2} - {k^\alpha u_\beta +u^\alpha k_\beta \over
k^\gamma u_\gamma}. 
\end{equation}
The $h^\alpha_\beta$ satisfy the relation
\begin{equation}
  \label{eq:opt8}
h^\alpha_\beta h^\beta_\gamma = h^\alpha_\beta, \ h^\alpha_\beta u^\beta 
= h^\alpha_\beta k^\beta = 0 \ {\rm and} \ h^\alpha_\alpha = 2.
\end{equation}
If $\bar{\delta} x^\alpha$ is the vector obtained by 
parallel-transporting 
$\delta_\bot x^\alpha$ from P$_{\rm s}$ along the ray, we have
\begin{equation}
  \label{eq:opt9}
\bar{\delta} x^\alpha = \delta_\bot x^\alpha + \int (\delta_\bot
x^\alpha)^\cdot dv,
\end{equation}
and the connection vector $\bar{\delta}_\bot x^\alpha$ vertical to
$u^\alpha$ and $k^\alpha$  at the point P$_{\rm o}$ is
\begin{equation}
  \label{eq:opt10}
\bar{\delta x^\alpha} = h^\alpha_\beta \bar{\delta} x^\beta =
\delta_\bot x^\alpha + \int  h^\alpha_\beta
(\delta_\bot x^\beta)^\cdot dv.
\end{equation}
The length $\delta l$ and the angle $\alpha_{12}$ are expressed as
follows using the connection vector in the plane vertical to
$k^\alpha$ and the screen velocity $\bar{u}^\alpha$ :
\begin{equation}
  \label{eq:opt11}
\delta l = (g_{\alpha\beta} \bar{\delta}x^\alpha
\bar{\delta}x^\beta)^{1/2}, \ \cos \alpha_{12} =
[(\bar{\delta}x^\alpha)_1 (\bar{\delta}x^\beta)_2]/[(\delta l)_1 
(\delta l)_2].
\end{equation}
Here $\bar{u}^\alpha$ is parallel-transformed along the ray from
$u^\alpha$ at P$_{\rm s}$.

\subsection{Optical scalars}
Using Eqs.(\ref{eq:opt4}) and (\ref{eq:opt5}) we obtain
\begin{equation}
  \label{eq:opt12}
(\delta_\bot x^\alpha)^\cdot = A^\alpha_\beta \delta_\bot x^\beta,
\end{equation}
where
\begin{equation}
  \label{eq:opt13}
A_{\alpha \beta} = h^\gamma_\alpha h^\lambda_\beta k_{\gamma ; \lambda}. 
\end{equation}
The tensor $A_{\alpha \beta}$ can be uniquely split as
\begin{eqnarray}
 \label{eq:opt14}
A_{\alpha \beta} &=& A_{[\alpha \beta]} + \theta h_{\alpha \beta} + 
\sigma_{\alpha \beta}, \cr
\omega &=& \Bigl[{1 \over 2}A_{[\alpha \beta]} A^{\alpha \beta}\Bigr]^{1 \over
2} = \Bigl[{1 \over 2}k_{[\alpha ;\beta]} k^{\alpha ;\beta}\Bigr]^{1 \over
2}, \cr
\theta &=& {1 \over 2} A^\gamma_\gamma = {1 \over 2} k^\gamma_{;\gamma}, 
\quad
\sigma_{\alpha \beta} = A_{(\alpha \beta)} - {1 \over
2}A^\gamma_\gamma h_{\alpha \beta}, \cr 
\sigma &=& \Bigl({1 \over 2} \sigma_{\alpha \beta} \sigma^{\alpha 
\beta}\Bigr)^{1 \over 2} = \Bigl\{{1 \over 2}[k_{(\alpha \beta)} 
k^{\alpha \beta}
- {1 \over 2}(k^\gamma_{;\gamma})^2]\Bigr\}^{1 \over 2},
\end{eqnarray}
where $\theta, \sigma$ and $\omega$ are optical scalars representing 
the expansion, shear and rotation, respectively, of ray bundles.
In geometric optics which we assume in the following, the rotation 
vanishes, because $k^\mu$ is a gradient vector.  By the transformation 
of $v$, $\theta, \sigma$ and $\omega$ transform, but $\theta dv,
\sigma dv$ and $\omega dv$ are invariant (Jordan et al.,\cite{rf:jord} 
Sachs\cite{rf:sachs61}).

The evolution equations for $\theta$ and $\sigma$ are
\begin{equation}
  \label{eq:opt15}
{d\theta \over dv} = {1 \over 2} k^\mu_{;\mu\nu} k^\nu  = -{1 \over 2}
{\cal R} - (\theta^2 + \sigma^2), 
\end{equation}
\begin{equation}
  \label{eq:opt16}
{d\sigma \over dv} = - {\cal C} -2 \theta \sigma,
\end{equation}
where ${\cal R} \equiv R_{\mu\nu} k^\mu k^\nu$, and ${\cal C}$ is
expressed in terms of the Weyl tensor as ${\cal C} \equiv
C_{\alpha\beta\gamma\lambda} k^\alpha k^{\gamma}\bar{t}^\beta
\bar{t}^{\lambda}$, with $t^\alpha$  a complex null vector satisfying 
$t^\alpha t_\alpha = k^\alpha t_\alpha = 0, \bar{t}^\alpha
\bar{t}_\alpha = 1$. As can be seen from Eqs.~(\ref{eq:opt15}) and
(\ref{eq:opt16}) there two terms causing the focusing of ray
bundles. One is the Ricci focusing term ${\cal R}$, proportional to the
matter density, and the other is the Weyl focusing term ${\cal C}$,
connected with the shear.

\subsection{Definition of distances}
 
The length of the shadow of the interval between two rays in the
observer plane is given by 
\begin{equation}
  \label{eq:opt17}
(dl)^2 = g_{\alpha \beta} 
\bar{\delta}_\bot x^\alpha \bar{\delta}_\bot x^\beta. 
\end{equation}
Then we have
\begin{equation}
  \label{eq:opt18}
{d (\delta l)\over \delta l} = A_{\alpha \beta} e^\alpha e^\beta dv = (\theta +
\sigma_{\alpha \beta} e^\alpha e^\beta) dv,
\end{equation}
where $e^\alpha \equiv \delta_\bot x^\alpha/ \delta l$ and $g_{\alpha
\beta} e^\alpha e^\beta = 1$. Next, let us consider the area of the
cross-section of a ray bundle given by
\begin{equation}
  \label{eq:opt19}
\delta A = {1 \over 2} {\int_0}^{2\pi} (\delta l)^2 d\alpha.
\end{equation}
If the deviation of the cross-section from a circle is small, we
obtain 
\begin{equation}
  \label{eq:opt20}
d(\delta A) = (\delta A)^\cdot dv = dv \delta A {\int_0}^{2\pi}
{d(\delta l) \over \delta  l} d\alpha
\end{equation}
or
\begin{equation}
  \label{eq:opt21}
{d(\delta A) \over \delta A} = 2 \theta dv,
\end{equation}
because ${\int_0}^{2\pi}\sigma_{\alpha \beta} e^\alpha e^\beta
d\alpha$ vanishes. 

Here we define two kinds of angular diameter distances (the linear 
angular diameter distance $D_{\rm lA}$ and the area angular diameter
distance  $D_{\rm aA}$) proportional to $\delta l$ and $(\delta A)^{1/2}$,
respectively, as
\begin{equation}
  \label{eq:opt22}
{d D_{\rm lA} \over dv} = (\theta + \sigma_{\alpha \beta} e^\alpha
e^\beta) D_{\rm lA},
\end{equation}
\begin{equation}
  \label{eq:opt23}
{d D_{\rm aA} \over dv} = \theta D_{\rm aA}, 
\end{equation}
where we consider ray bundles with $\theta = \infty$ at the
observer point P$_{\rm o}$. In the case of no shear, $D_{\rm lA}$ 
and $D_{\rm aA}$ 
are equal, but generally they are different. Their average values are
equal if the term $\sigma_{\alpha \beta} e^\alpha e^\beta$ is
cancelled out in the averaging process. In the situation that 
$\theta= \infty$ at the source's point P$_{\rm s}$, we obtain the luminosity 
distance $D_{\rm L}$ satisfying
\begin{equation}
  \label{eq:opt24}
{d D_{\rm L} \over dv} = \theta D_{\rm L}.
\end{equation}
The relation between $D_{\rm aA}$ and $D_{\rm L}$ is proved to be 
\begin{equation}
  \label{eq:opt25}
D_{\rm L} = (1 + z)^2 D_{\rm aA}
\end{equation}
by Etheringen,\cite{rf:ether} where $z$ is the redshift 
given by the 
relation $1 + z = (u^\alpha k_\alpha)_{\rm source} \\
/(u^\alpha k_\alpha)_{\rm observer}$.

The solutions of Eqs. (\ref{eq:opt22}) and (\ref{eq:opt23})
are generally complicated, but they can easily be obtained in the
special case in which (1) the density is spatially constant, (2) there is
no shear, and (3) the affine parameter in the Friedman background can
be used. From Eqs. (\ref{eq:opt15}) and (\ref{eq:opt23}) we obtain
\begin{equation}
  \label{eq:opt26}
{d^2 D_{\rm A} \over dv^2} = -{1 \over 2} {\cal R} D_{\rm A},
\end{equation}
where $D_{\rm A} = D_{\rm lA} = D_{\rm aA}$, ${\cal R} = 8\pi G \alpha \rho/ a^2 (t) 
= 3 (1+z)^5 \alpha \Omega_0$,  $\Omega_0$ is the total density
parameter, and $\alpha$ is the smoothing (clumpiness) parameter. 
The affine parameter $v$ is related to $z$ as
\begin{equation}
  \label{eq:opt27}
{dz \over dv} = (1 +z)^2 [(1+\Omega_0 z)(1+ z)^2 - \lambda_0 z(2+z)]^{1/2},
\end{equation}
where $\lambda_0$ is the normalized cosmological constant. The
boundary condition for $D_{\rm A}$ at epoch $z = z_1$ is
\begin{equation}
  \label{eq:opt28}
D_{\rm A} = 0, \quad {d D_{\rm A} \over dz} = c (dt/dz)_{z=z_1} = {c \over
H_0}{a(z_1) \over a_0}. 
\end{equation}
The solution of Eq. (\ref{eq:opt27}) is called the Friedmann
 distance for $\alpha = 1$, the Dyer-Roeder distance for
$\alpha = 0$, and the generalized Dyer-Roeder distance for arbitrary
$\alpha$.

For the analyses of cosmological lens systems, we often use the lens 
equation
\begin{equation}
  \label{eq:opt29}
\mbox{\boldmath $\beta$} = \mbox{\boldmath $\theta$} - 
4G{D_{OL} D_{LS} \over D_{OS}} 
{\partial \over \partial \mbox{\boldmath $\theta$}}
\int d^2 \theta' \Sigma (\theta') \ln {\vert \theta -\theta'\vert 
\over \theta_c},
\end{equation}
where {\boldmath $\beta$} and {\boldmath $\theta$} are 
the angular position vectors (as seen by the observer) of the image 
and source, respectively, relative to the lens, $\Sigma (\theta')$ is 
the surface mass density of the lens on the lens plane, $\theta_c$ 
is an arbitrary constant angle,
and $D_{OL}, D_{OS}$ and $D_{LS}$ are the angular
distances between the lens and the observer, the source and the
observer, and the lens and the source, respectively. This equation 
has so far been derived only from intuitive geometrical considerations 
with use of the 
thin lens approximation, but it is not clear what distances should be 
used. In order to derive the lens equation from cosmological equations 
of light propagation, Sasaki\cite{rf:sasaki93} used the equation of 
geodesic deviation.
It is obtained from Eq. (\ref{eq:opt4}) for the connection vector:
\begin{equation}
  \label{eq:opt30}
(\delta x^\mu)^{\cdot \cdot} = - R^\mu_{\nu\alpha\beta} k^\nu
k^\beta \delta x^\alpha.
\end{equation}
He solved this equation in the case of an ideal light path, which 
is separated into three
regions: the homogeneous, shearless region I (between the observer and
the lens object), the region II including the lens object, and the
homogeneous, shearless region III (between the lens object and the
source). The light path in regions I and III is expressed by the
generalized Dyer-Roeder distance, and the deflection of light rays in
region III is determined using the thin lens approximation. The resulting 
lens equation reduces to the usual one with the generalized 
Dyer-Roeder distances. In the case that in the regions I and III there
are inhomogeneous matter distributions and the shear effect is not
negligible, however, the usual expression of the lens equation cannot 
be used, as was shown by Sasaki.\cite{rf:sasaki93}

In the multi-lens-plane method,\cite{rf:sef}\cite{rf:sw88a}\tocite{rf:sw88b}
 inhomogeneities as lens objects are
assumed to be only in the lens planes, and hence the use of the lens
equation in this method is consistent with the above assumption that
the regions I and III are homogeneous and shearless.  
However the neglection of gravitational forces due to the difference 
of the projected matter distribution from the real distribution may be 
comparable with the neglection of weak forces from distant matter 
distribution.

\bigskip
\section{Distances and lensing relations}

\subsection{Observation and distances in gravitational lensing} 

There are some methods to determine cosmological parameters 
by using gravitational lenses. 
\cite{Refsdal64a,Refsdal64b,Refsdal66,PG,BN,FFK,rf:fuk92,rf:sef} 
Most of them concern the following three typical
observational quantities: 
(1) the bending angle, (2) the lensing statistics and (3) the time delay. 
It is of great importance to clarify the determination of cosmological 
parameters through their observations in the realistic universe. 
In particular, it has been discussed that inhomogeneities of 
the universe may affect the cosmological tests. 
\cite{rf:fs89,rf:zel64,rf:dash66,rf:dyroe72,rf:dyroe73,rf:wts92,rf:sasaki93,rf:sw88a,rf:sw88b,Kantowski,Linder88,KFT,BS91}

In this section, we use the so-called Dyer-Roeder angular diameter 
distance in order to take account of the inhomogeneities. 
\cite{rf:dyroe72,rf:dyroe73,DR74}
We can consider this distance in two different cases, that of
the so-called filled beam in the Friedmann-Lemaitre-Robertson-Walker (FLRW) 
universe, and that of the 
so-called empty beam, when the right ray propagates through the empty region. 
For comparison with the filled beam, the empty beam has been frequently 
used and studied numerically in the literature (for instance, Fukugita et al. 
\cite{FFK,rf:fuk92}).
However, it has not been clarified whether the observed quantities and/or 
the cosmological parameters for the arbitrary case of 
the clumpiness parameter are bounded between those for the filled beam 
and the empty beam. 
Moreover, numerical investigations have fixed redshifts of the lens and 
the source, \cite{AA,rf:fuk92} though the effect of the clumpiness 
on the observable depends on the redshifts of the lens and the source. 
Therefore, it is important to clarify how the observation of gravitational 
lensing depends on all the parameters (the density parameter, 
cosmological constant, clumpiness parameter 
and redshifts of the lens and the source).  For this reason, 
we derive the dependence on these parameters. \cite{Asada97,rf:asad98} 

\subsection{Distance combinations in gravitational lenses}

\noindent
(1) bending angle

The lens equation is written as \cite{rf:sef}   
\begin{equation}
\mbox{\boldmath $\beta$}=\mbox{\boldmath $\theta$}-{D_{\rm LS} \over D_{\rm OS}}
\mbox{\boldmath $\alpha$} . 
\label{angle}
\end{equation}
Here, {\boldmath $\beta$} and {\boldmath $\theta$} are 
the angular position vectors of the source and image, respectively, 
and {\boldmath $\alpha$} is the vector representing the deflection angle. 
The effective bending angle $(D_{\rm LS} / D_{\rm OS})\mbox{\boldmath $\alpha$}$ 
appears when we discuss the observations concerning the angle such as 
the image separation and the location of the critical line. 
\cite{BN,rf:sef}
Hence the ratio $D_{\rm LS} / D_{\rm OS}$ plays an important role in the discussion 
of observations concerning the angle. 
It has been argued that, in calculating the bending angle, the density 
along the line of sight should be subtracted from the density of 
the lens object. \cite{rf:sasaki93} 
However, we assume that the density of the lens is much larger than 
that along the line of sight, so that the effect of the clumpiness 
on {\boldmath $\alpha$} can be ignored. 
Thus, we consider only the ratio $D_{\rm LS} / D_{\rm OS}$ in the following. 

\noindent
(2) lensing statistics 

The differential probability of lensing events is \cite{PG,rf:sef} 
\begin{equation}
d\tau=\sigma n_{\rm L} dl , 
\label{statistics}
\end{equation}
where 
$n_{\rm L}$ is the number density of the lens, $dl$ is the physical 
length of the 
depth and $\sigma$ is the cross section, proportional to 
$D_{\rm OL}D_{\rm LS} / D_{\rm OS}$. 
Since $dl$ depends only on the cosmological parameters in the FLRW universe, 
we investigate the combination $D_{\rm OL}D_{\rm LS} / D_{\rm OS}$ in order to 
take account of the clumpiness of the matter. 

\noindent
(3) time delay

The time delay between two images A and B is written as 
\cite{Refsdal64b,rf:sef} 
\begin{equation}
\Delta t_{\rm AB}={1+z_L \over c}{D_{\rm OL}D_{\rm OS} \over D_{\rm LS}}
\int^{\rm B}_{\rm A} d\mbox{\boldmath $\theta$}\cdot\Bigl(
{ \mbox{\boldmath $\alpha$}_{\rm A}+\mbox{\boldmath $\alpha$}_{\rm B} \over 2 }
-\mbox{\boldmath $\alpha$}(\mbox{\boldmath $\theta$}) \Bigr) , 
\label{delay}
\end{equation}
where {$\mbox{\boldmath $\alpha$}_{\rm A}$} and {$\mbox{\boldmath 
$\alpha$}_{\rm B}$} 
are the bending angles at the images A and B, respectively.

\subsection{Monotonic properties}

It is assumed that the affine parameter in the Dyer-Roeder distance 
is the same as that in the FLRW universe,\cite{rf:dyroe72,rf:sasaki93} 
namely Eq.~$(\ref{eq:opt27})$. 
Since $\alpha$ represents the strength of the Ricci focusing 
along the line of sight, the DR angular diameter distance is 
a decreasing function of $\alpha$ for a fixed redshift,\cite{rf:dyroe73} 
that is to say,
\begin{equation}
D_{\rm OL}(\alpha_1) > D_{\rm OL}(\alpha_2) \quad\mbox{for}\quad 
\alpha_1 < \alpha_2 . 
\label{drtime1}
\end{equation}

\noindent
(1) $D_{\rm LS}/D_{\rm OS}$

It has been shown that the distance ratio $D_{\rm LS}/D_{\rm OS}$ satisfies 
\cite{Asada97} 
\begin{equation}
{ D_{\rm LS} \over D_{\rm OS} }(\alpha_1) < 
{ D_{\rm LS} \over D_{\rm OS} }(\alpha_2) \quad\mbox{for}\quad \alpha_1 < \alpha_2 . 
\label{drbend1}
\end{equation}
This is shown as follows. 
For fixed $z_{\rm S}$, $\Omega_0$ and $\lambda_0$, the ratio $D_{\rm
LS}/D_{\rm OS}$ 
can be considered as a function of $z_{\rm L}$, $X_{\alpha}(z_{\rm L})$. 
We define $Y_{\alpha}(z_{\rm L})$ as $D_{\rm SL}/D_{\rm OS}$, 
where $D_{\rm SL}$ is the Dyer-Roeder distance from the source to the lens. 
Owing to the reciprocity, \cite{rf:ether}  we obtain 
\begin{equation}
Y_{\alpha}(z_{\rm L})={1+z_{\rm S} \over 1+z_{\rm L}}X_{\alpha}(z_{\rm L}) . 
\label{reciprocity}
\end{equation}
Since $Y_{\alpha}$ depends on $z_{\rm L}$ only through $D_{\rm SL}$, it obeys
the equation  
\begin{equation}
{d^2 \over d{v_{\rm L}}^2}Y_{\alpha}(z_{\rm L})+{3 \over 2}(1+z_{\rm L})^5 
\alpha \Omega_0 Y_{\alpha}(z_{\rm L})=0 , \label{raychaudhuri2}
\end{equation}
where $v_{\rm L}$ is an affine parameter at the lens. 
Let us define the Wronskian as 
\begin{equation}
W(Y_{\alpha_1},Y_{\alpha_2})=Y_{\alpha_1}{d Y_{\alpha_2} \over 
d v_{\rm L}}-Y_{\alpha_2}{d Y_{\alpha_1} \over d v_{\rm L}} . 
\end{equation}
Then, we obtain 
\begin{equation}
{d \over d v_{\rm L}}W(Y_{\alpha_1},Y_{\alpha_2}) < 0 
\quad\mbox{for}\quad \alpha_1 < \alpha_2 . 
\label{wronskian2}
\end{equation}
Since both $Y_{\alpha_1}$ and $Y_{\alpha_2}$ vanish at $z_L=z_{\rm S}$, 
we obtain 
\begin{equation}
W(Y_{\alpha_1}(z_{\rm S}),Y_{\alpha_2}(z_{\rm S}))=0 . \label{wronskian3} 
\end{equation}
{}From Eqs.~$(\ref{wronskian2})$ and $(\ref{wronskian3})$, we find 
\begin{equation}
W(Y_{\alpha_1},Y_{\alpha_2}) >0 , \label{wronskian4}
\end{equation}
where we used the fact that the affine parameter $v$ defined 
by Eq.~$(\ref{eq:opt27})$ is an increasing function of $z$. 
Equation ($\ref{wronskian4}$) can be rewritten as 
\begin{equation}
{d \over d v_{\rm L}} \ln{Y_{\alpha_2} \over Y_{\alpha_1}} > 0 . 
\label{wronskian5}
\end{equation}
Since $Y_{\alpha}$ always becomes $1+z_{\rm S}$ at the observer, we find 
\begin{equation}
\ln{Y_{\alpha_2}(z_{\rm L}=0) \over Y_{\alpha_1}(z_{\rm L}=0)}=0 . 
\label{wronskian6}
\end{equation}
{}From Eqs.~$(\ref{wronskian5})$ and $(\ref{wronskian6})$, we obtain 
\begin{equation}
{Y_{\alpha_2} \over Y_{\alpha_1}}>1 . \label{proof2}
\end{equation}
Thus, from Eq.~$(\ref{reciprocity})$, Eq.~$(\ref{drbend1})$ is proved. 

{}From Eqs.~$(\ref{angle})$ and $(\ref{drbend1})$, we see the image separation 
as well as the effective bending angle {\it increases} with $\alpha$. 

\noindent
(2) $D_{\rm OL}D_{\rm LS} / D_{\rm OS}$

Next let us prove that $D_{\rm OL}D_{\rm LS} / D_{\rm OS}$ increases monotonically 
with $\alpha$. 
We fix $\Omega$, $\lambda$, $z_{\rm L}$ and $z_{\rm S}$. 
Then it is crucial to note that the distance from the lens 
to the source can be expressed in terms of the distance function from 
the observer, $D(z)$, as \cite{Linder88}
\begin{equation}
D_{\rm LS}={c \over H_0} (1+z_{\rm L}) D_{\rm OL}D_{\rm OS} \int^{v_{\rm S}
}_{v_{\rm L}} {dv \over D(z)^2} ,  
\label{dls}
\end{equation}
where $H_0$ is the Hubble constant at present. 
This can be rewritten as 
\begin{equation}
{D_{\rm OL}D_{\rm LS} \over D_{\rm OS}}(\alpha)={c \over H_0}(1+z_{\rm L}) 
D_{\rm OL}\,^2 \int^{v_{\rm S}}_{v_{\rm L}} {dv \over D(z)^2} . 
\label{drstat1} 
\end{equation}
The right hand side of this equation depends on $\alpha$ only through 
$D_{\rm OL} / D(z)$. 
Following reasoning similar to that used in the proof of 
Eq.~$(\ref{drbend1})$, 
we obtain for $z_{\rm L} < z  < z_{\rm S}$ 
\begin{equation}
{D_{\rm OL} \over D(z)}(\alpha_1) < {D_{\rm OL} \over D(z)}(\alpha_2) 
\quad\mbox{for}\quad \alpha_1 < \alpha_2 . 
\label{drstat2}
\end{equation}
{}From Eqs.~$(\ref{drstat1})$ and $(\ref{drstat2})$, we obtain 
\begin{equation}
{D_{\rm OL}D_{\rm LS} \over D_{\rm OS}}(\alpha_1) < 
{D_{\rm OL}D_{\rm LS} \over D_{\rm OS}}(\alpha_2) 
\quad\mbox{for}\quad \alpha_1 < \alpha_2 . 
\label{drstat3}
\end{equation}
Therefore, the gravitational lensing event rate {\it increases} 
with $\alpha$. 

\noindent
(3) $D_{\rm LS} / D_{\rm OL}D_{\rm OS}$

Finally, we investigate the combination of distances appearing 
in the time delay. 
Dividing Eq.~$(\ref{drtime1})$ by Eq. $(\ref{drbend1})$, we obtain 
\begin{equation}
{D_{\rm OL}D_{\rm OS} \over D_{\rm LS}}(\alpha_1) >
{D_{\rm OL}D_{\rm OS} \over D_{\rm LS}}(\alpha_2) 
\quad\mbox{for}\quad \alpha_1 < \alpha_2 .
\label{drtime2}
\end{equation}
Thus, the time delay {\it decreases} with $\alpha$. 

As shown above, the three types of combinations of distances 
are monotonic functions of the clumpiness parameter. 
However, some of other combinations of distances are not monotonic 
functions of $\alpha$, though these combinations may not be necessarily 
related with the observation. 
For instance, the combination $D_{\rm LS} / \sqrt{c D_{\rm OS} / H_0}$ 
is not a monotonic function of $\alpha$. 

\subsection{Implications for cosmological tests}

We consider three types of the cosmological test which use 
combinations of distances appearing in gravitational lensing. 
Let us fix the density parameter in order to discuss constraints on 
the cosmological constant. 

\noindent
(1) $D_{\rm LS}/D_{\rm OS}$

The following relation holds 
\begin{equation}
{D_{\rm LS} \over D_{\rm OS}}(\lambda_1) < {D_{\rm LS} \over 
D_{\rm OS}}(\lambda_2) \quad\mbox{for} \quad \lambda_1 < \lambda_2 . 
\label{lambend1}
\end{equation}
This is shown as follows. 
Let us define 
\begin{equation}
X_{\lambda}(z_{\rm L})={D_{\rm LS}(\alpha,\Omega_0,\lambda) \over 
D_{\rm OS}(\alpha,\Omega_0,\lambda)}  
\end{equation}
and 
\begin{equation}
Y_{\lambda}(z_{\rm L})={D_{\rm SL}(\alpha,\Omega_0,\lambda) \over 
D_{\rm OS}(\alpha,\Omega_0,\lambda)} . 
\end{equation}
By the reciprocity,\cite{rf:ether} we obtain 
\begin{equation}
Y_{\lambda}(z_{\rm L})={1+z_{\rm S} \over 1+z_{\rm L}}X_{\lambda}(z_{\rm L}) , 
\label{reciprocity2}
\end{equation}
which satisfies 
\begin{equation}
{d^2 \over {dv_{\rm L}}^2}Y_{\lambda}(z_{\rm L})+{3 \over 2}
(1+z_{\rm L})^5 \alpha \Omega_0 
Y_{\alpha}(z_{\rm L})=0 . 
\label{raychaudhuri21}
\end{equation}
For $\lambda_i\,(i=1,2)$, the affine parameter $v_i$ satisfies 
\begin{equation}
{dz_{\rm L} \over dv_i}=(1+z_{\rm L})^2 \sqrt{ \Omega_0 z_{\rm L} 
(1+z_{\rm L})^2 -\lambda_i 
z_{\rm L}(2+z_{\rm L})+(1+z_{\rm L})^2 } . 
\label{affine21}
\end{equation}
We define the Wronskian as 
\begin{equation}
W(Y_{\lambda_1},Y_{\lambda_2})=Y_{\lambda_1}{d Y_{\lambda_2} \over 
d v_2}-Y_{\lambda_2}{d Y_{\lambda_1} \over d v_1} . 
\label{wronskian21}
\end{equation}
Then, using Eq.~$(\ref{raychaudhuri21})$, we obtain 
\begin{equation}
{d \over d z_{\rm L}}W(Y_{\lambda_1},Y_{\lambda_2}) < 0 
\quad\mbox{for}\quad \lambda_1 < \lambda_2 . 
\label{wronskian22}
\end{equation}
Since $Y_{\lambda}$ always vanishes at $z_{\rm L}=z_{\rm S}$, we also obtain 
\begin{equation}
W(Y_{\lambda_1}(z_{\rm S}),Y_{\lambda_2}(z_{\rm S}))=0 . 
\label{wronskian23} 
\end{equation}
{}From Eqs.~$(\ref{wronskian22})$ and $(\ref{wronskian23})$, we find 
\begin{equation}
W(Y_{\lambda_1},Y_{\lambda_2}) >0 
\quad\mbox{for}\quad \lambda_1 < \lambda_2 , 
\label{wronskian24}
\end{equation}
which can be rewritten as 
\begin{equation}
{d \over d z_{\rm L}} \ln{Y_{\lambda_2} \over Y_{\lambda_1}} > 0 
\quad\mbox{for}\quad \lambda_1 < \lambda_2 . 
\label{wronskian25}
\end{equation}
Since $Y_{\lambda}$ always becomes $1+z_{\rm S}$ at the observer, we have
\begin{equation}
\ln{Y_{\lambda_2}(z_{\rm L}=0) \over Y_{\lambda_1}(z_{\rm L}=0)}=0 . 
\label{wronskian26}
\end{equation}
{}Finally from Eqs.~$(\ref{wronskian25})$ and $(\ref{wronskian26})$, we obtain 
\begin{equation}
{Y_{\lambda_2} \over Y_{\lambda_1}} >1 
\quad\mbox{for}\quad \lambda_1 < \lambda_2 . 
\label{proof21}
\end{equation}
Thus, Eq.~$(\ref{lambend1})$ is proved. 

Equations $(\ref{drbend1})$ and $(\ref{lambend1})$ imply that, 
in a cosmological test using the bending angle, the cosmological constant 
estimated by use of the distance formula in the FLRW universe is always 
{\it less} than that etimated by use of the Dyer-Roeder distance 
$(0 \leq \alpha <1)$. 

\noindent
(2) $D_{\rm OL}D_{\rm LS}/D_{\rm OS}$

Multiplying Eq. $(\ref{lambend1})$ by
\begin{equation}
D_{\rm OL}(\lambda_1) < D_{\rm OL}(\lambda_2) 
\quad\mbox{for}\quad \lambda_1 < \lambda_2 , 
\label{lamstat1}
\end{equation}
we obtain 
\begin{equation}
{D_{\rm OL}D_{\rm LS} \over D_{\rm OS}}(\lambda_1) < 
{D_{\rm OL}D_{\rm LS} \over D_{\rm OS}}(\lambda_2) 
\quad\mbox{for}\quad \lambda_1 < \lambda_2 . 
\label{lamstat2}
\end{equation}
Equation $(\ref{lamstat1})$ can be proved, for instance,
in the following manner: 
The Dyer-Roeder distance is written as the integral equation 
\cite{rf:sw88a,Linder88}  
\begin{equation}
D(z;\alpha)=D(z;\alpha=1)+\sum^{\infty}_{i=1} \left[ {3 \over 2} 
{c \over H_0} (1-\alpha) \Omega \right]^i \int^z_0 dy K_i(y,z) 
D(y;\alpha=1) , 
\label{DRlam2}
\end{equation}
where $K_i(y,z)$ is defined as 
\begin{equation}
K_1(x,y)=\left.{dv \over dz}\right|_{z=x} (1+x)^4 D(x,y;\alpha=1) 
\label{K1}
\end{equation}
and 
\begin{equation}
K_{i+1}(x,y)=\int^y_x dz K_1(x,z)K_i(z,y) . 
\label{Ki}
\end{equation}
{}From Eqs. $(\ref{K1})$ and $(\ref{Ki})$, 
it is shown that for $x<y$ 
\begin{equation}
K_i(x,y;\lambda_1) < K_i(x,y;\lambda_2) 
\quad \mbox{for} \quad \lambda_1 < \lambda_2 , 
\label{Kilam}
\end{equation}
where we have used the relation
\begin{equation}
D(x,y;\alpha=1,\lambda_1) < D(x,y;\alpha=1,\lambda_2) 
\quad \mbox{for} \quad \lambda_1 < \lambda_2 , 
\label{FLRWlam} 
\end{equation}
applicable in the FLRW universe. Using Eqs. $(\ref{DRlam2})$, 
$(\ref{Kilam})$ and $(\ref{FLRWlam})$, 
and the positivity of $K_i$, we obtain Eq. $(\ref{lamstat1})$. 

{}From Eqs. $(\ref{drstat3})$ and $(\ref{lamstat2})$, it is found that, 
in a cosmological test using the lensing events rate, 
the cosmological constant is always {\it underestimated} 
by use of the distance formula in the FLRW universe.

\noindent
(3) $D_{\rm LS}/D_{\rm OL}D_{\rm OS}$
 
When the time delay is measured and the lens object is observed, 
$D_{\rm OL}D_{\rm OS} / D_{\rm LS}$ can be determined from 
Eq. ($\ref{delay}$). 
On the other hand, when we denote the dimensionless distance between 
$z_1$ and $z_2$ as $d_{12}=H_0 D_{12} / c$, which does not depend on 
the Hubble constant, we obtain 
\begin{equation}
{D_{\rm OL}D_{\rm OS} \over D_{\rm LS}}={c \over H_0} 
{d_{\rm OL}d_{\rm OS} \over d_{\rm LS}} . 
\label{delay2}
\end{equation}
Then, Eq. ($\ref{drtime2}$) becomes 
\begin{equation}
{d_{\rm OL}d_{\rm OS} \over d_{\rm LS}}(\alpha_1) > {d_{\rm OL}d_{\rm OS} \over d_{\rm LS}} 
(\alpha_2) \quad\mbox{for}\quad \alpha_1 < \alpha_2 . 
\label{drtime3}
\end{equation}
Thus, from Eqs.($\ref{delay2}$) and ($\ref{drtime3}$), it is found that 
$H_0$ estimated using the Dyer-Roeder distance {\it decreases} 
with $\alpha$. 
Thus, the Hubble constant can be bounded from below when we have little 
knowledge on the clumpiness of the universe. 
The lower bound is given by use of the distance 
in the FLRW universe. 
On the other hand, since the combination $D_{\rm LS} / D_{\rm OL}D_{\rm OS}$ 
is {\it not} a monotonic function of the cosmological constant, 
the relation between the clumpiness of the universe and 
the cosmological constant is not simple. 

It should be noted that even the assumption of a spatially flat 
universe $(\Omega+\lambda=1)$ does not change the above implications 
for the three types of cosmological tests, since 
the cosmological constant affects the Dyer-Roeder distance formula 
only through the relation between $z$ and $v$, Eq. $(\ref{eq:opt27})$. 

\subsection{Evolution of clumpiness} 

We have taken the clumpiness parameter $\alpha$ as a constant 
along the line of sight. 
However, as a reasonable extension of the DR distance, $\alpha$ can be 
considered as a function of the redshift in order to take account of 
the growth of inhomogeneities of the universe.\cite{Linder88} 
In proving the monotonic properties, it has never been assumed that 
$\alpha$ is constant. 
Hence, all the monotonic properties and the implications for cosmological 
tests remain unchanged for the variable $\alpha(z)$. 
That is to say, when $\alpha_1(z) < \alpha_2(z)$ is always satisfied 
for $0<z<z_{\rm S}$, all we must to do is to replace parameters $\alpha_1$ and 
$\alpha_2$ with functions $\alpha_1(z)$ and $\alpha_2(z)$ in 
Eqs. $(\ref{drbend1})$, $(\ref{drstat3})$ and $(\ref{drtime2})$.  
In particular, when $\alpha(z)$ is always less than unity on the 
way from the source to the observer, both of the combinations of 
distances appearing in (1) and (2) are less, while the combination 
in (3) is larger than those in the FLRW universe.
Then, the decrease in the bending angle and the lensing event rate, 
and the increase in the time delay hold even for a generalized DR 
distance with variable $\alpha(z)$. 

\newpage

\section{Average distances and the dispersions in inhomogeneous models}
In this section we describe the statistical behavior of angular diameter 
distances in 
inhomogeneous model universes at the stage of $0 < z < z_1 (= 5)$.
To derive the distances we use the light rays received by (or emitted 
backwards from)
an observer at present (by solving the null-geodesic equation) in the
 universes which were produced numerically.

\subsection{Model universes and lens models}

We consider three background models with 
$(\Omega_0, \lambda_0) = (1.0, 0), (0.2, 0.8)$ and $(0.2, 0)$. 
They are denoted as S, L and O, respectively, which represent the
standard model, a low-density flat model and an open model. The matter is 
assumed to contain
particles consisting of galaxies and non-galactic clouds with equal
mass $m$, but generally different sizes. The inhomogeneous models are
given by the method of the $N$-body simulation (using Suto's tree 
code)\cite{rf:su}
in periodic boxes with  particle number $N = 32^3$. 
The initial particle distributions were derived using Bertschinger's
software {\it COSMICS}\cite{rf:bert}
 under the condition that their perturbations are given
as random fields with the spectrum of cold dark matter, their power
$n$ is 1, their normalization is specified as the dispersion 
$\sigma_8 \ = \ 0.94$, and the Hubble constant is $H_0 = 100 h {\rm
Mpc^{-1} \ km \ s^{-1}}$, where $h = 0.5$ for $(1.0,0)$ and 
$h = 0.7$ for other models with $\Omega_0 = 0.2$.

The box sizes for models S, L and O are 
\begin{equation}
  \label{eq:ba1}
L_0 \equiv a(t_0) l = 32.5 h^{-1}, \ 50 h^{-1}, \ 50 h^{-1}{\rm Mpc},
\end{equation}
and the particle masses are 
\begin{equation}
  \label{eq:ba2}
m (= \rho_{B0} {L_0}^3/N) = 2.90, \ 2.11, \ 2.11 \times 10^{11} h^{-1} 
M_\odot,
\end{equation}
respectively, where $\rho_{B0}$ is the background mass density, $a(t)$
is the scale-factor, and $l$ is the comoving length.

The particle size $r_s \ (= a(t) x_s)$ is given in the form of  
softening
radii, which have constant values when we calculate the gravitational
potential for lensing. For $r_s$ we consider the following two (lens)
models:

\noindent \mib{Lens \ model \ 1}. \quad
 All particles in the low-density models ($\Omega_0 =
0.2$) have $r_s  = 20 h^{-1}$kpc, $20 \%$ of the particles in
the flat model $(1.0, 0)$ have $r_s = 20 h^{-1}$kpc, and the
remaining particles have  $r_s = 500 h^{-1}$kpc.
Thus, practically, particles with $\Omega_c = 0.2$ (which we call {\it compact 
lens objects}) play the role of lens objects. Their number density is
much larger than the galactic density $\Omega_g \sim 0.02$. 

\noindent \mib{Lens \ model \ 2}. \quad
$10 \%$ of the particles in the low-density model ($\Omega_0 = 0.2$)
and $2 \%$ of the particles in the flat model $(1.0, 0)$ have 
$r_s = 20 h^{-1}$kpc, while the remaining particles have 
 $r_s = 500 h^{-1}$kpc. Therefore only galaxies corresponding to
$\Omega_g = 0.02$ play significant roles as lens objects, and the remaining
particles are regarded as diffuse clouds.

The background line-element is
\begin{equation}
  \label{eq:ba3}
ds^2 = - c^2 dt^2 + a^2(t)(d\mib{x})^2/\Bigl[1 + K {1 \over 4}
(\mib{x})^2\Bigr]^2,
\end{equation}
and the Poisson equation and null-geodesic equation describing light rays
are given in \S 2 of a separate paper (by Tomita, Premadi and 
Nakamura) of this volume.

\subsection{Angular diameter distances}

Here we treat the linear and area distances defined in \S 2
(Eqs. ($\ref{eq:opt22}$) and ($\ref{eq:opt23}$)).
Let us consider a pair of rays received by the observer with the
separation angle $\theta$. By solving null-geodesic equations,
 the interval of the two rays at any epoch can be derived.
If $(\Delta \mib{x})_\perp$ is the component of the deviation vector
perpendicular to the central direction of the rays, the linear angular 
diameter distance $D_{\rm lA}$ is given as
\begin{equation}
  \label{eq:ba4}
D_{\rm lA} = a(t) (\Delta \mib{x})_\perp \Bigl[1 + {1 \over 4} K
(\mib{x})^2\Bigr]^{-1}/\theta,
\end{equation}
where the factor $(1-2\Phi)$  has been neglected, because
$\vert \Phi \vert \ll 1$ locally. The above expression can be
rewritten by use of $y^i \ (\equiv a_0 x^i/R_0)$ as
\begin{equation}
  \label{eq:ba5}
D_{\rm lA} = {R_0 \over (1+z) F} (\Delta \mib{y})_\perp/\theta,
\end{equation}
where $F \equiv 1 -{1 \over 4}(R_0H_0/c)^2 (1 -\Omega_0 -\lambda_0) 
(\mib{y})^2, \ a_0 = a(t_0) $ and \ $R_0 \equiv L_0/N^{1/3}$.

On the other hand, the area angular diameter distance $D_{\rm aA}$ is
given as follows using three rays (ray 1, ray 2 and ray 3) received by
the observer, such that on the observer plane the two lines between
ray 1 and ray 2, and between ray 1 and ray 3 are orthogonal and have
the same lengths (equal to the separation angle $\theta$).
If $(\Delta \mib{x})_{\perp(12)}, \ (\Delta \mib{x})_{\perp(13)}$  and
$(\Delta \mib{x})_{\perp(23)}$ are the components of the deviation vectors
(between ray 1 and ray 2, \ between ray 1 and ray 3 \ and between 
ray 2 and ray 3 ) perpendicular to the central direction of 
the rays, we obtain
\begin{equation}
  \label{eq:ba6}
D_{\rm aA} = a(t) [(\Delta \mib{x})_{\perp(12)} \cdot (\Delta
\mib{x})_{\perp(13)}]^{1/2} \Bigl[1 + {1 \over 4} K (\mib{x})^2
\Bigr]^{-1}/\theta.
\end{equation}
Using $y^i \ (\equiv a_0 x^i/R_0)$, this reduces to
\begin{equation}
  \label{eq:ba7}
D_{\rm aA} = {R_0 \over (1+z) F} \Big[{1 \over 2} \Delta y_{12} \Delta
y_{13} |(\Delta y_{12} - \Delta y_{13} + \Delta y_{23})(\Delta y_{12}
- \Delta y_{13} - \Delta y_{23})|\Big]^{1/4}/\theta, 
\end{equation}
where $\Delta y_p \equiv |(\Delta \mib{y})_{\perp (p)}|$ with
$p = 12, 13, 23$.

In a previous paper (Tomita\cite{rf:tom98c}) we investigated the behavior of 
$D_{\rm lA}$ for the
separation angle $\theta = 0.005 - 20$ arcsec in various model
universes, and found the dependence of distances on $\theta$ is small for 
$\theta \leq 1.0$ arcsec. Here we fix the separation angle to $\theta
= 1.0$ arcsec and 
consider the difference between the linear and area distances and
their dependence on the lens models 1 and 2.

In the present lensing simulation we performed the ray-shooting of 500 
ray bundles to derive $D_{\rm lA}$ and $D_{\rm aA}$ for each set of two lens 
models and three model universes. At the six epochs 
$z = 0.5, 1, 2, 3, 4$ and $5$, we compared the calculated distances
with the Dyer-Roeder distance and determined the corresponding value 
of the clumpiness parameter $\alpha$ as follows.
In Ref. 17) we calculated $\alpha$ for $0 \leq z \leq 5$ in the above
three model universes and found that the angular diameter distance 
depends approximately linearly on $\alpha$ \ (cf. Figs. 3 $\sim$ 6 in
Ref. 17). For $|\alpha -1| \gg 1$  linearity does not hold
(cf. Eq. (3.35)), but most light rays are in the neighborhood of
$\alpha = 1$, as is verified below. Hence we determined $\alpha$ 
from the calculated distance 
$D_{\rm A}~(= D_{\rm lA}$ or $D_{\rm aA})$ using the relation
\begin{equation}
  \label{eq:d6}
\alpha = (D_{\rm A} - D_{\rm DR})/(D_{\rm F} - D_{\rm DR}),
\end{equation}
where $D_{\rm DR}$ is the limiting Dyer-Roeder distance with $\alpha = 0$,
and $D_{\rm F}$ is the calculated Friedmann distance in the homogeneous
case. This $D_{\rm F}$ is equal to the Dyer-Roeder distance with 
$\alpha = 1$. Moreover, we consider the normalized distances defined by
\begin{equation}
  \label{eq:d7}
d_{\rm A} = D_{\rm A}/D_{\rm F}. 
\end{equation}

\begin{figure}[htb]
 \parbox{\halftext}{
\epsfxsize=7.5cm
\epsfbox{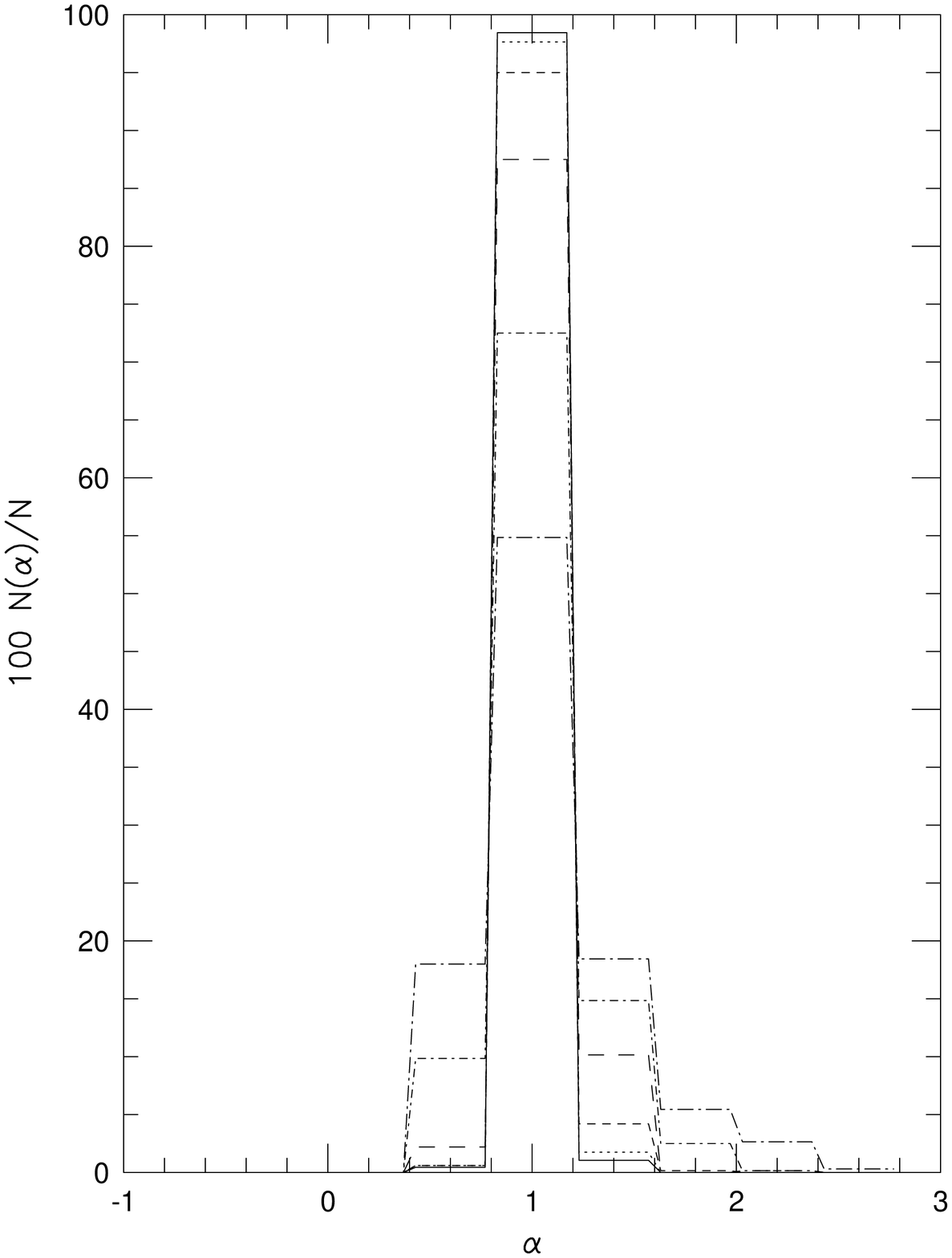}
   \caption{The percentage ($100 N(\alpha)/ N)$ of the distribution 
of $\alpha$ in bins with the interval $\Delta \alpha = 0.4$, for 
$D_{\rm lA}$ in the lens model 1 and model S with $(\Omega_0, \lambda_0) =
(1.0, 0)$. Results for $z = 0.5, 1, 2, 3, 4$ and $5$ are denoted
by dot-long dashed, dot-short dashed, long dashed, short dashed, 
dotted and solid lines, respectively.}}
\hspace{-2mm}
 \parbox{\halftext}{
\epsfxsize=7.5cm
\epsfbox{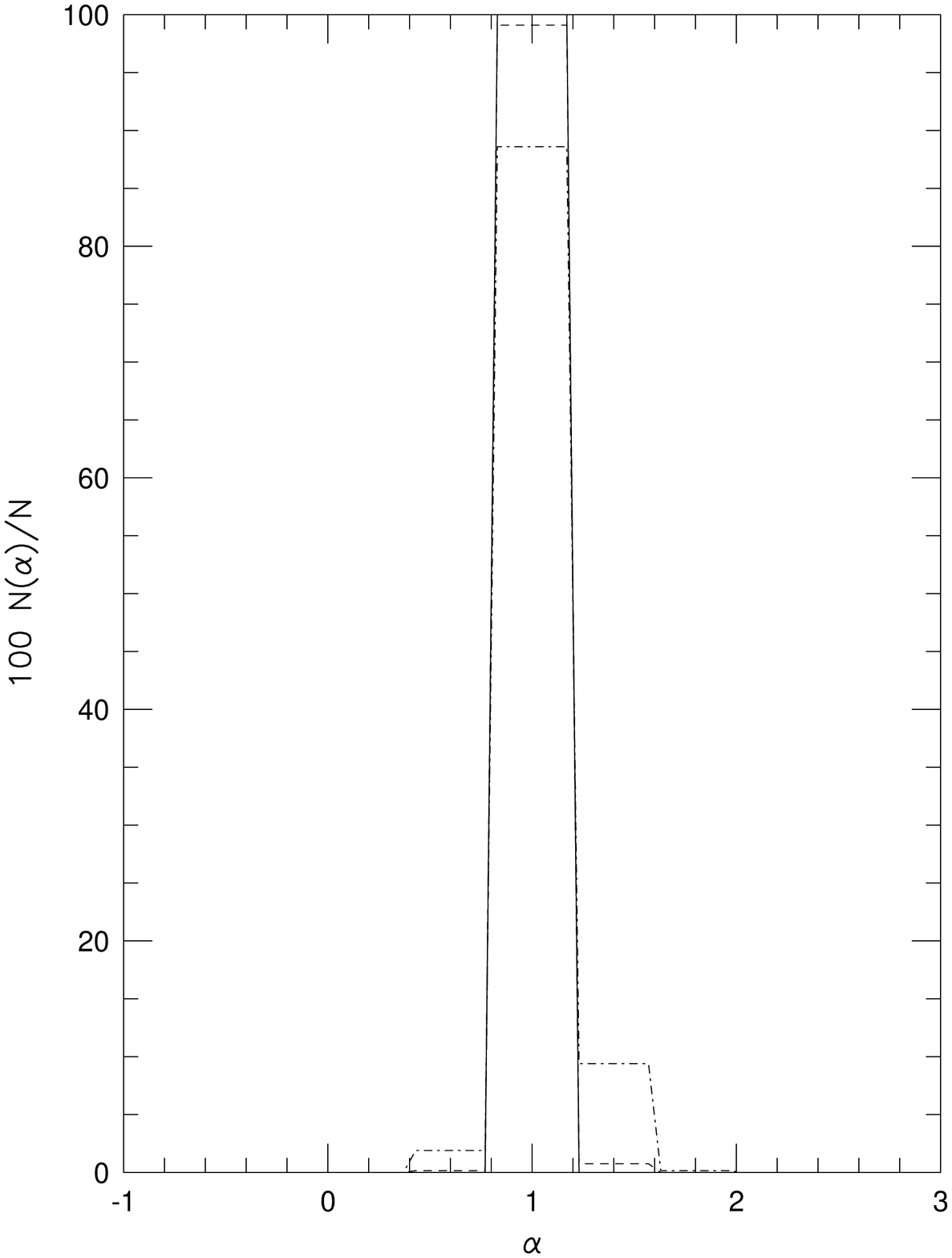}
   \caption{The percentage ($100 N(\alpha)/ N)$ of the distribution 
of $\alpha$ in bins with the interval $\Delta \alpha = 0.4$, for 
$D_{\rm lA}$ in the lens model 2 and model S with $(1.0, 0)$.
  The lines have the same meaning as in Fig. 1. 
}}
\end{figure}

\begin{figure}[htb]
 \parbox{\halftext}{
\epsfxsize=7.5cm
\epsfbox{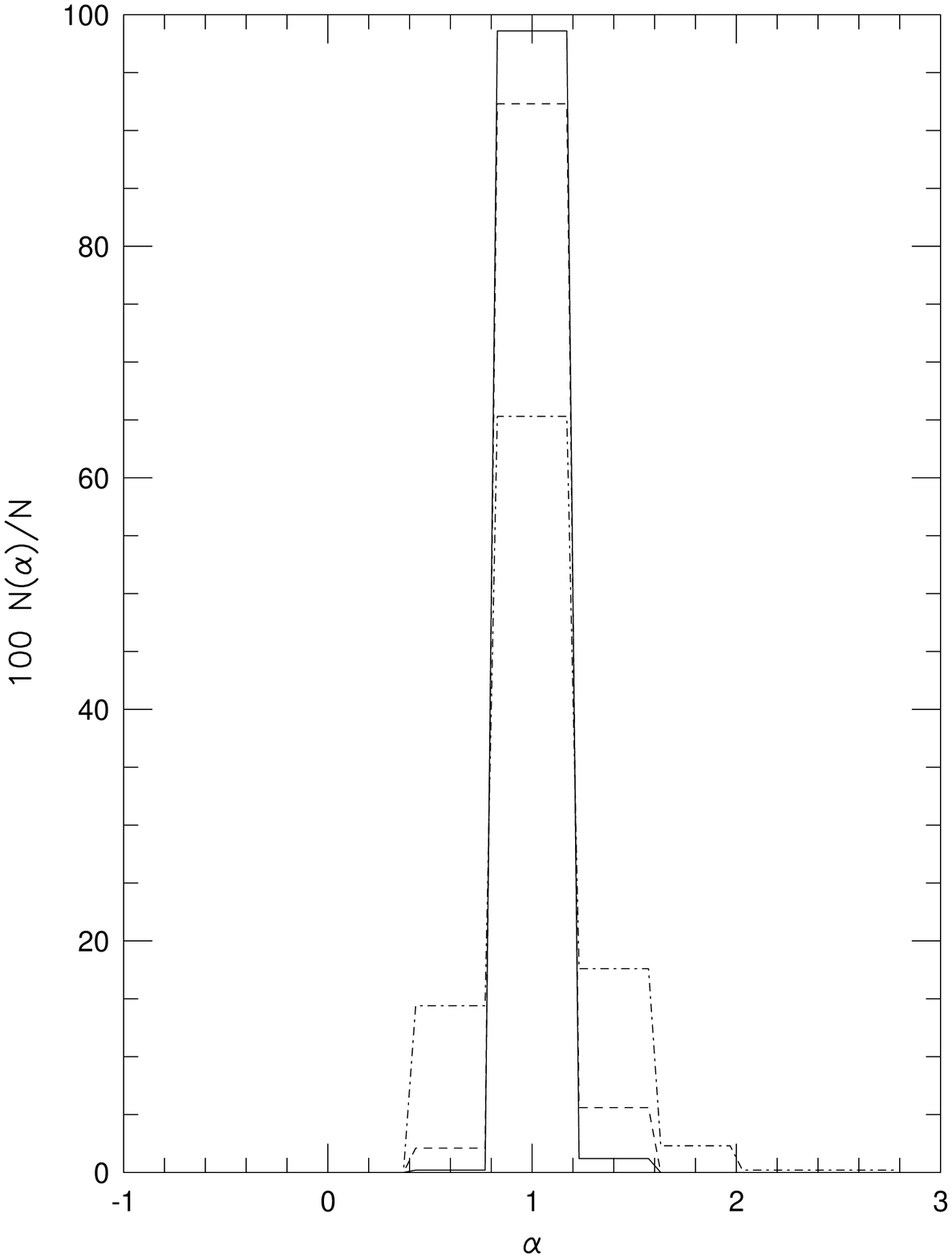}
   \caption{The percentage ($100 N(\alpha)/ N)$ of the distribution 
of $\alpha$ in bins with the interval $\Delta \alpha = 0.4$, for 
$D_{\rm aA}$ in the lens model 1 and model S with $(1.0, 0)$. 
The lines have the same meaning as in Fig. 1. }}
\hspace{-2mm}
 \parbox{\halftext}{
\epsfxsize=7.5cm
\epsfbox{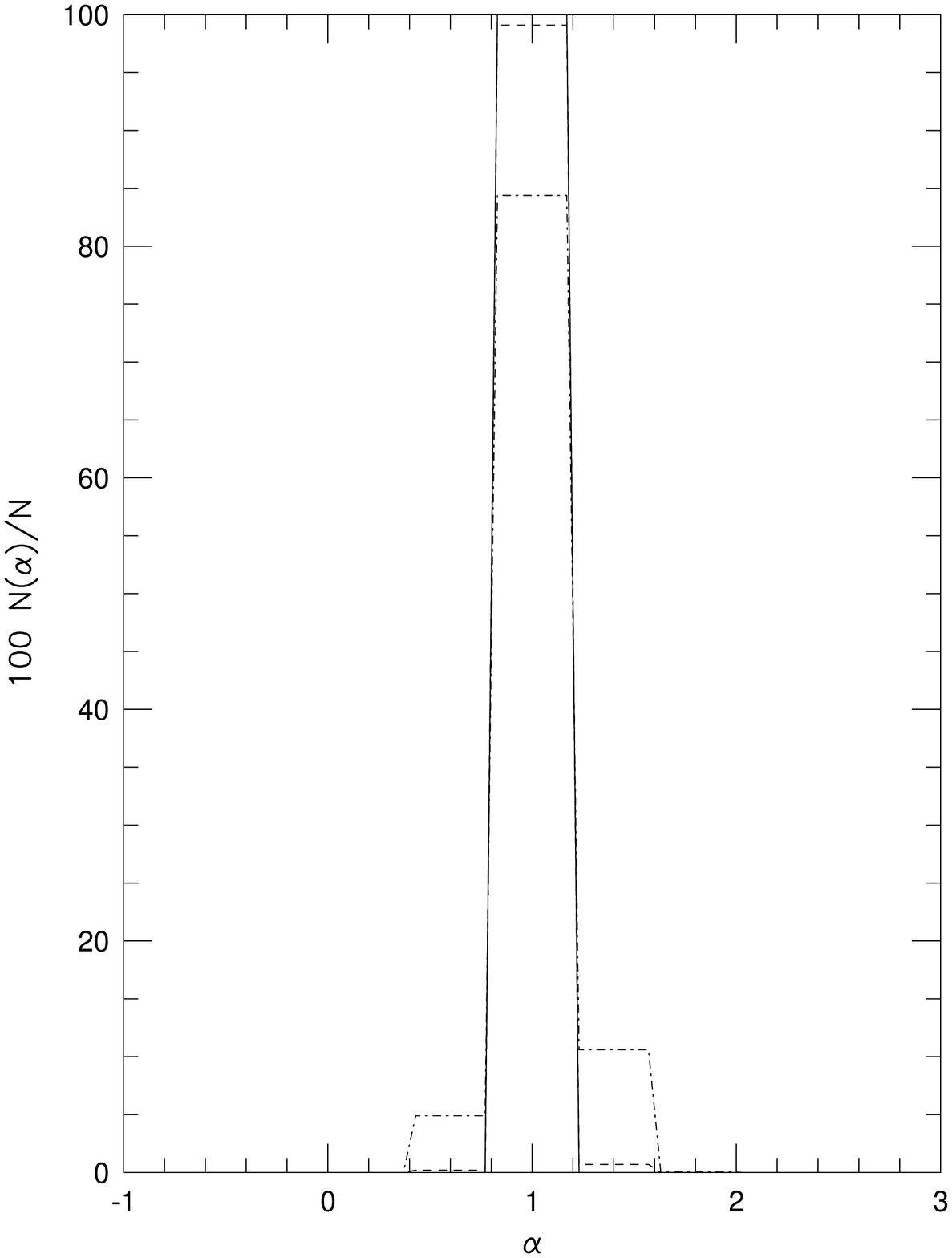}
   \caption{The percentage ($100 N(\alpha)/ N)$ of the distribution 
of $\alpha$ in bins with the interval $\Delta \alpha = 0.4$, for 
$D_{\rm aA}$ in the lens model 2 and model S with (1.0, 0).
  The lines have the same meaning as in Fig. 1. }}
\end{figure}
   
\begin{figure}[htb]
 \parbox{\halftext}{
\epsfxsize=7.5cm
\epsfbox{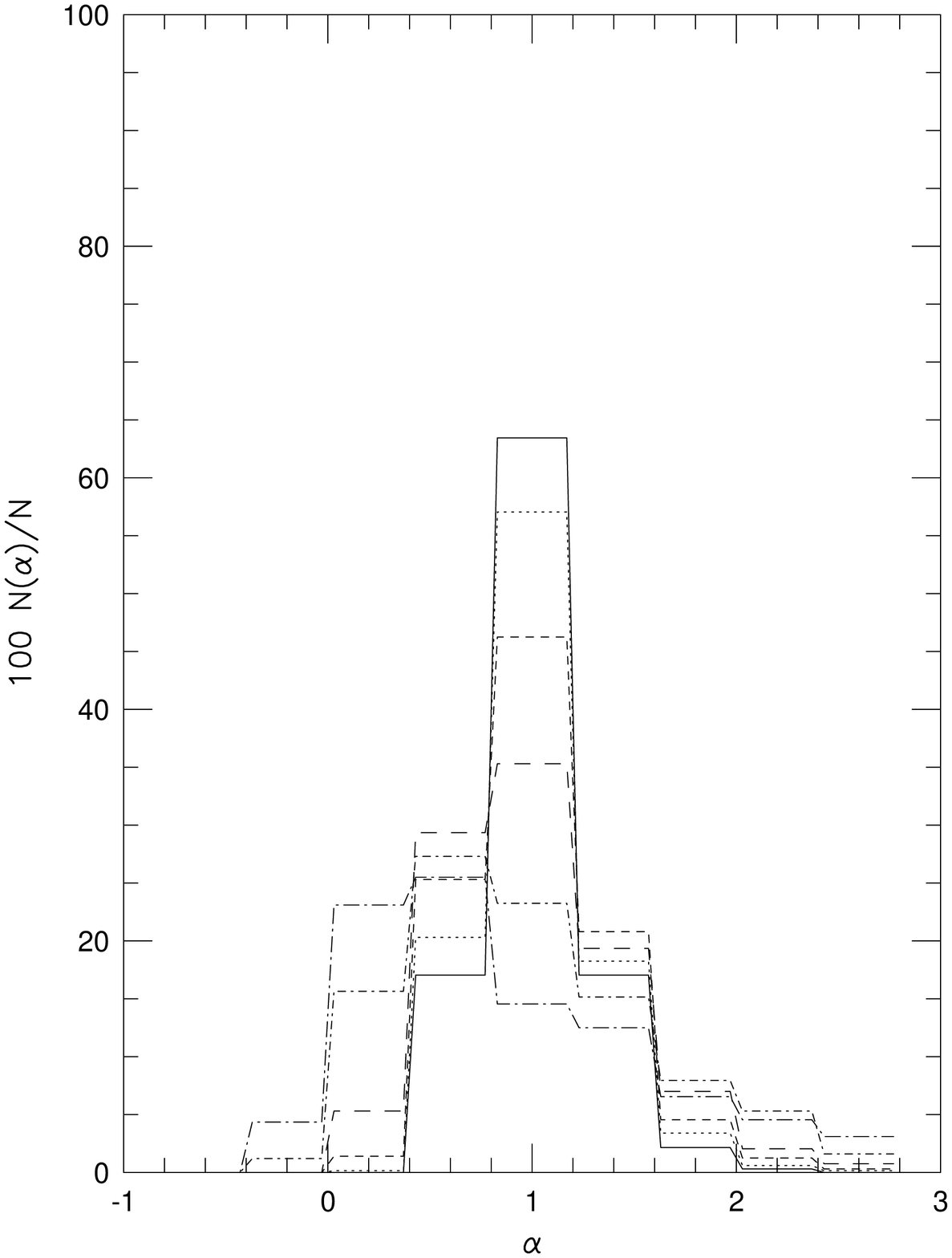}
   \caption{The percentage ($100 N(\alpha)/ N)$ of the distribution 
of $\alpha$ in bins with the interval $\Delta \alpha = 0.4$, for 
$D_{\rm lA}$ in the lens model 1 and model L with $(0.2, 0.8)$. 
The lines have the same meaning as in Fig. 1. }}
\hspace{-2mm}
 \parbox{\halftext}{
\epsfxsize=7.5cm
\epsfbox{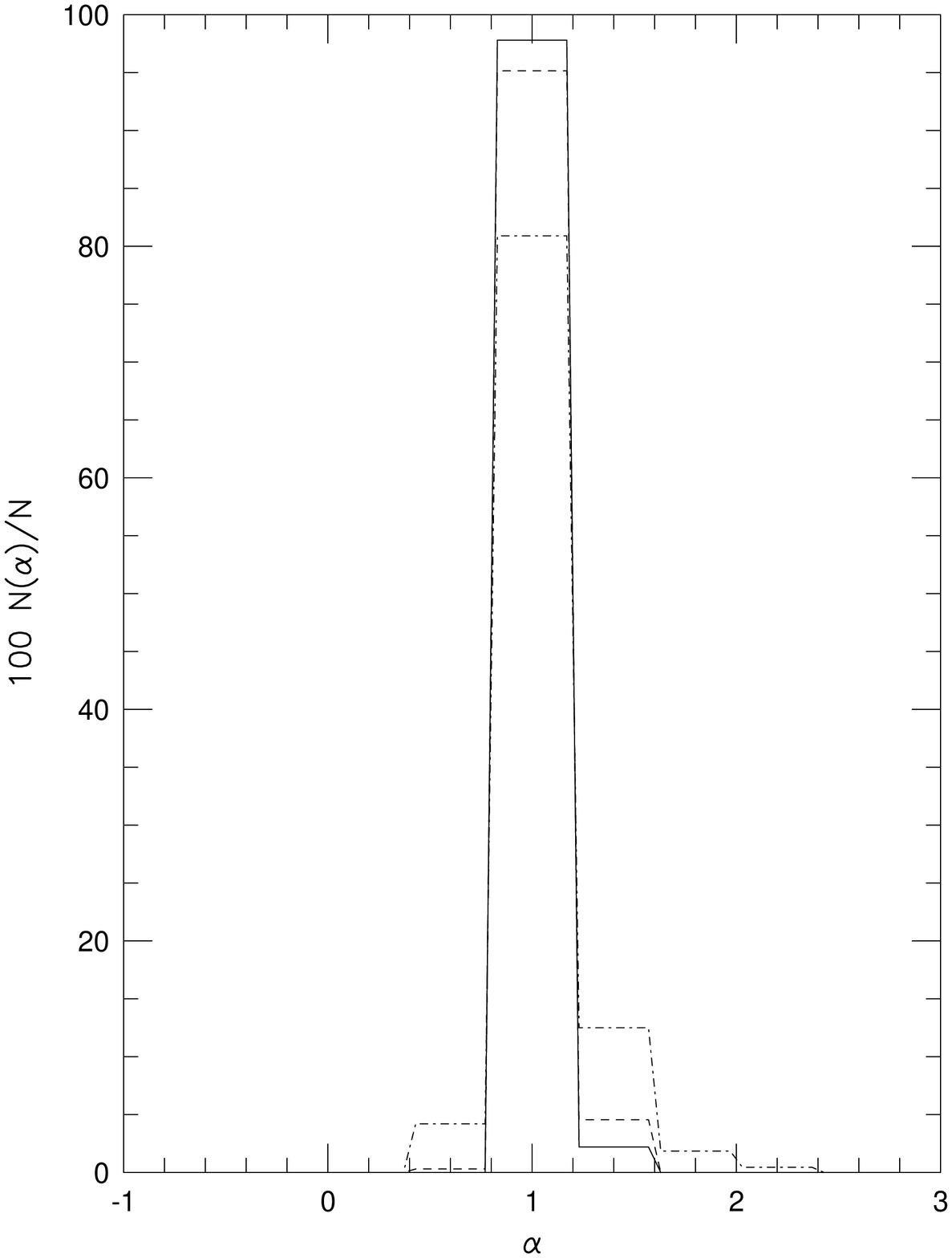}
   \caption{The percentage ($100 N(\alpha)/ N)$ of the distribution 
of $\alpha$ in bins with the interval $\Delta \alpha = 0.4$, for 
$D_{\rm lA}$ in the lens model 2 and model O1 with (0.2, 0.8).
  The lines have the same meaning as in Fig. 1. }}
\end{figure}

\begin{table}
\caption{The average clumpiness parameter $\bar{\alpha}$ and its dispersion 
$\sigma_\alpha$, and the average normalized distance $\bar{d}_{\rm A}$ 
and its dispersion $\sigma_d$ for two lens models
 in model S with $(\Omega_0, \lambda_0) = (1.0, 0)$. \
}

\label{table:1}
\begin{center}
\begin{tabular}{cccccccccc} \hline \hline
& & &linear & $(D_{\rm lA})$& & &area &$(D_{\rm aA})$ & \\ \hline
lens & $z$ &$\bar{\alpha} $ & $\sigma_\alpha$ & $\bar{d}_{\rm A}
 $ & $\sigma_d$&$\bar{\alpha} $ & $\sigma_\alpha$ & $\bar{d}_{\rm A}
 $ & $\sigma_d$
\\ \hline
 &0.5& $1.09$ &  $0.43$ &  1.00 &  0.018 &$1.09$ &  $0.36$ &  1.00 &  0.015\\
 &1&   $1.03$ &  $0.25$ &  1.00 &  0.032 &$1.04$ &  $0.22$ &  1.00 &  0.028\\
1&2&   $1.02$ &  $0.15$ &  0.99 &  0.051 &$1.03$ &  $0.41$ &  0.99 &  0.045\\
 &3&   $1.01$ &  $0.11$ &  0.99 &  0.063 &$1.02$ &  $0.10$ &  0.99 &  0.056\\
 &4&   $1.01$ &  $0.09$ &  0.99 &  0.073 &$1.02$ &  $0.08$ &  0.99 &  0.064\\
 &5&   $1.01$ &  $0.08$ &  0.99 &  0.080 &$1.02$ &  $0.07$ &  0.99 &  0.071\\
\hline
 &0.5& $1.09$ &  $0.28$ &  1.00 &  0.012 &$1.07$ &  $0.23$ &  1.00 &  0.009\\
 &1&   $1.03$ &  $0.14$ &  1.00 &  0.019 &$1.03$ &  $0.13$ &  1.00 &  0.017\\
2&2&   $1.02$ &  $0.09$ &  0.99 &  0.029 &$1.02$ &  $0.08$ &  0.99 &  0.027\\
 &3&   $1.01$ &  $0.07$ &  0.99 &  0.037 &$1.02$ &  $0.06$ &  0.99 &  0.034\\
 &4&   $1.01$ &  $0.06$ &  0.99 &  0.043 &$1.01$ &  $0.05$ &  0.99 &  0.040\\
 &5&   $1.01$ &  $0.05$ &  0.99 &  0.048 &$1.01$ &  $0.04$ &  0.99 &  0.044\\
\hline
\end{tabular}
\end{center}
\end{table}

\begin{table}
\caption{The average clumpiness parameter $\bar{\alpha}$ and its dispersion 
$\sigma_\alpha$, and the average normalized distance $\bar{d}_{\rm A}$ 
and its dispersion $\sigma_d$ for two lens models
 in model L with $(\Omega_0, \lambda_0) = (0.2, 0.8)$. \
}

\label{table:2}
\begin{center}
\begin{tabular}{cccccccccc} \hline \hline
& & &linear & $(D_{\rm lA})$ & & &area & $(D_{\rm aA})$ & \\ \hline
lens & $z$ &$\bar{\alpha} $ & $\sigma_\alpha$ & $\bar{d}_{\rm A}
 $ & $\sigma_d$&$\bar{\alpha} $ & $\sigma_\alpha$ & $\bar{d}_{\rm A}
 $ & $\sigma_d$
\\ \hline
 &0.5& $1.09$ &  $1.43$ &  1.00 &  0.019 &$1.02$ &  $0.91$ &  1.00 &  0.012\\
 &1&   $1.07$ &  $1.15$ &  1.00 &  0.057 &$1.02$ &  $0.69$ &  1.00 &  0.034\\
1&2&   $1.02$ &  $0.68$ &  1.00 &  0.114 &$1.03$ &  $0.49$ &  0.99 &  0.083\\
 &3&   $1.01$ &  $0.50$ &  1.00 &  0.157 &$1.03$ &  $0.37$ &  0.99 &  0.118\\
 &4&   $1.00$ &  $0.39$ &  1.00 &  0.188 &$1.02$ &  $0.29$ &  0.99 &  0.138\\
 &5&   $1.00$ &  $0.33$ &  1.00 &  0.211 &$1.02$ &  $0.24$ &  0.99 &  0.155\\
\hline
 &0.5& $1.08$ &  $0.43$ &  1.00 &  0.006 &$1.06$ &  $0.29$ &  1.00 &  0.004\\
 &1&   $1.04$ &  $0.33$ &  1.00 &  0.016 &$1.03$ &  $0.21$ &  1.00 &  0.010\\
2&2&   $1.01$ &  $0.20$ &  1.00 &  0.034 &$1.01$ &  $0.13$ &  1.00 &  0.022\\
 &3&   $1.01$ &  $0.15$ &  1.00 &  0.047 &$1.01$ &  $0.10$ &  1.00 &  0.032\\
 &4&   $1.01$ &  $0.12$ &  1.00 &  0.056 &$1.01$ &  $0.08$ &  1.00 &  0.039\\
 &5&   $1.01$ &  $0.10$ &  1.00 &  0.064 &$1.01$ &  $0.07$ &  0.99 &  0.045\\
\hline
\end{tabular}
\end{center}
\end{table}
\bigskip

\begin{figure}[htb]
 \parbox{\halftext}{
\epsfxsize=7.5cm
\epsfbox{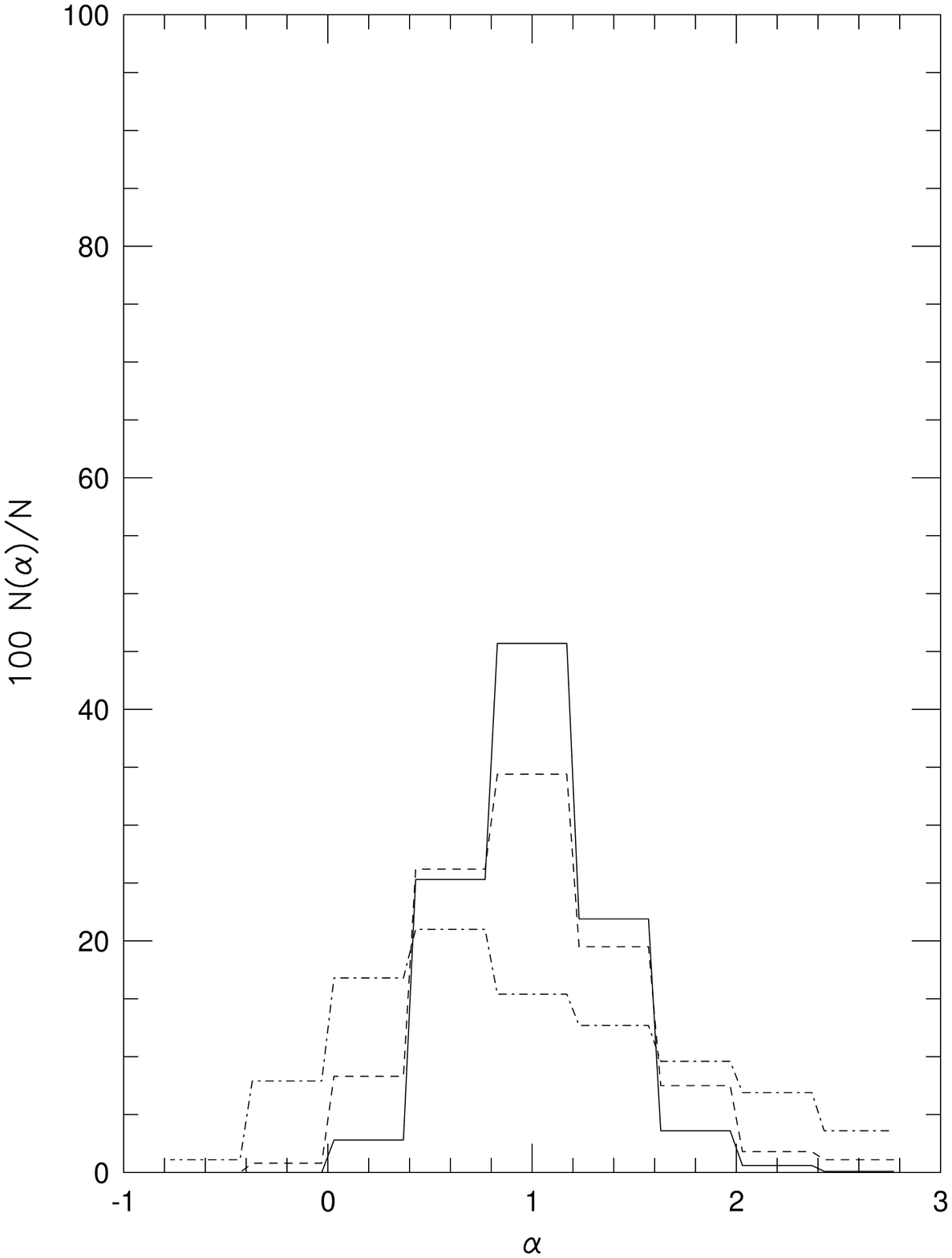}
   \caption{The percentage ($100 N(\alpha)/ N)$ of the distribution 
of $\alpha$ in bins with the interval $\Delta \alpha = 0.4$, for 
$D_{\rm aA}$ in the lens model 1 and model O2 with $(0.2, 0.8)$. 
The lines have the same meaning as in Fig. 1. }}
\hspace{-2mm}
 \parbox{\halftext}{
\epsfxsize=7.5cm
\epsfbox{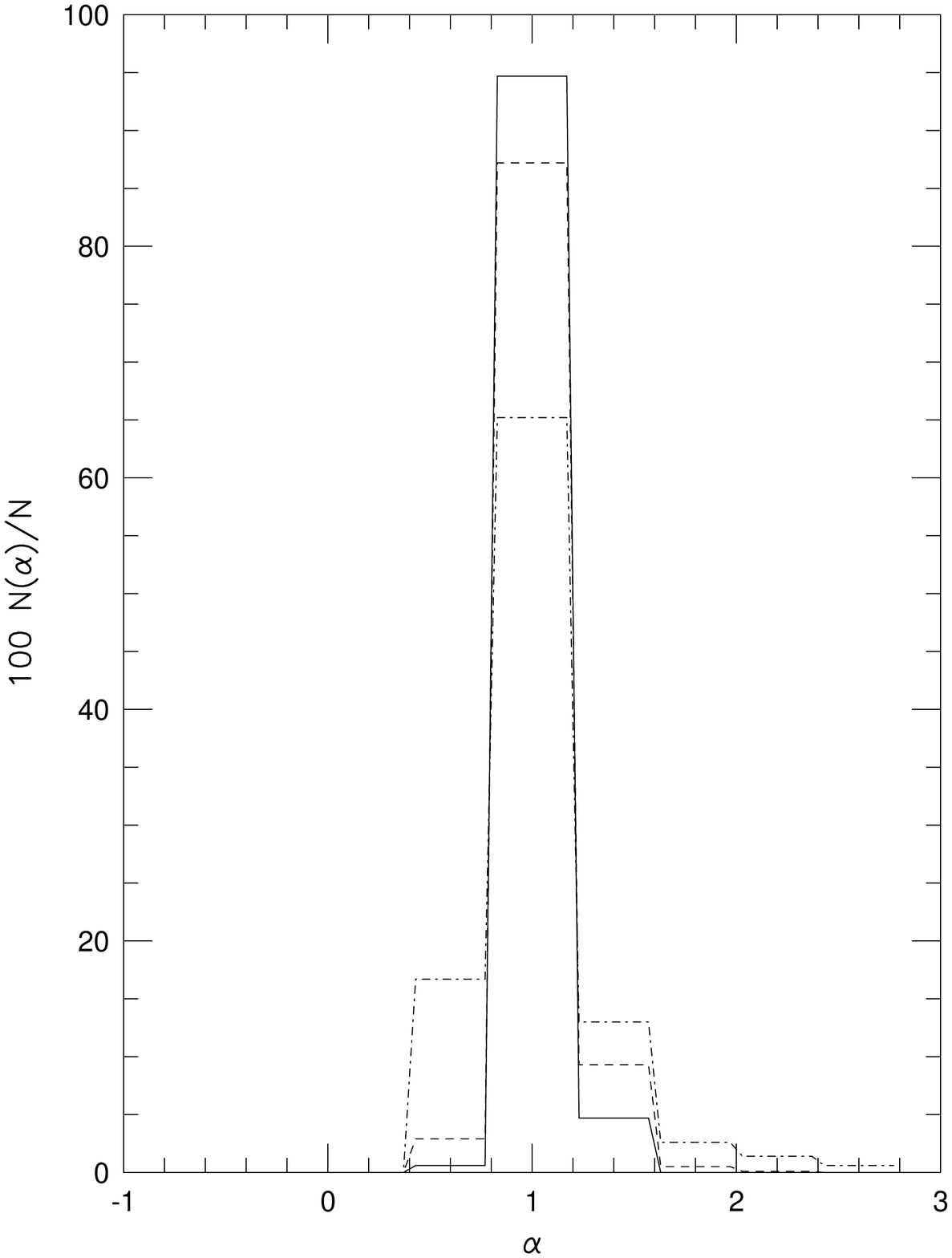}
   \caption{The percentage ($100 N(\alpha)/ N)$ of the distribution 
of $\alpha$ in bins with the interval $\Delta \alpha = 0.4$, for 
$D_{\rm aA}$ in the lens model 2 and model O2 with (0.2, 0.8).
  The lines have the same meaning as in Fig. 1. }}
\end{figure}

\begin{figure}[htb]
 \parbox{\halftext}{
\epsfxsize=7.5cm
\epsfbox{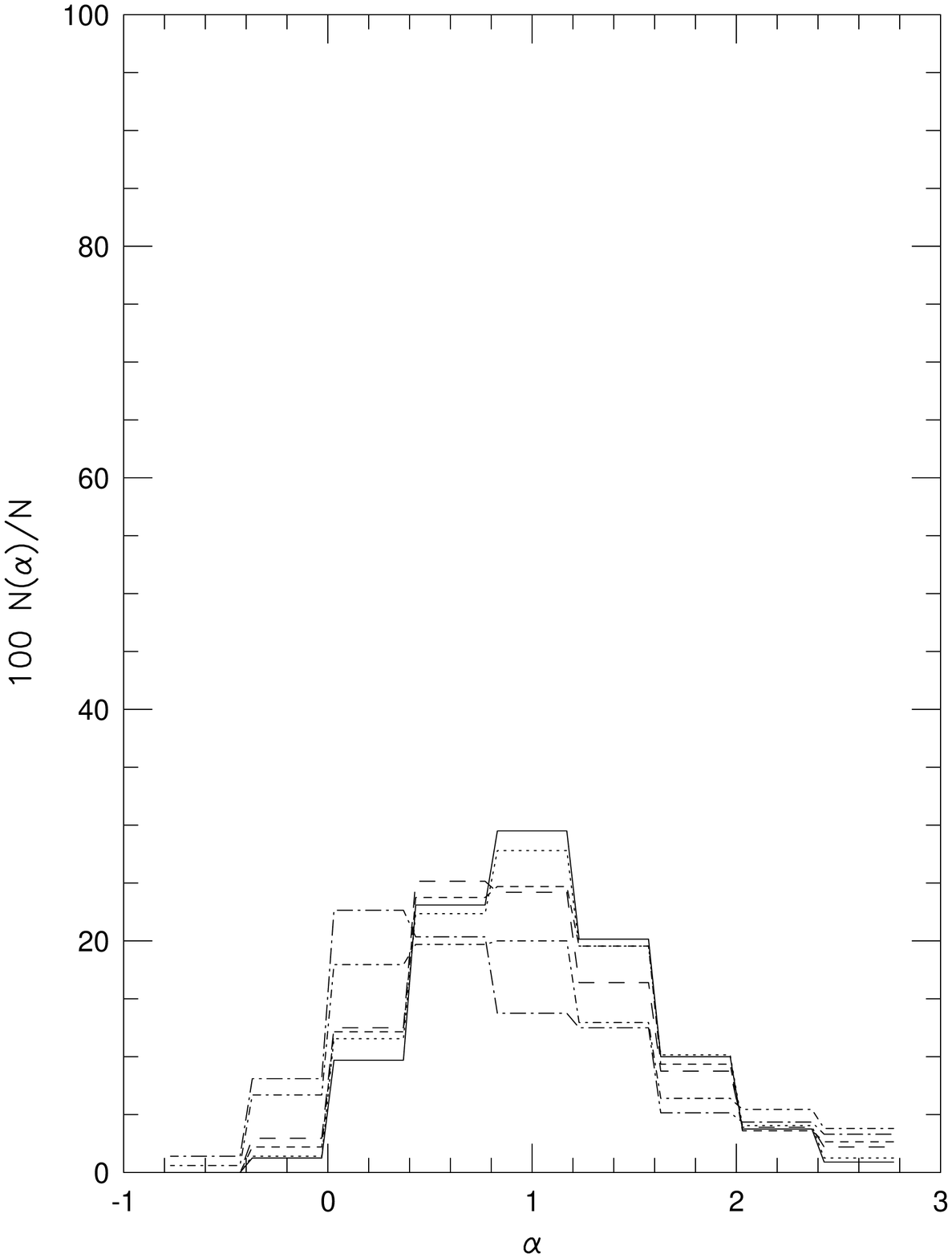}
   \caption{The percentage ($100 N(\alpha)/ N)$ of the distribution 
of $\alpha$ in bins with the interval $\Delta \alpha = 0.4$, for 
$D_{\rm lA}$ in the lens model 1 and model O with $(0.2, 0)$. 
The lines have the same meaning as in Fig. 1. }}
\hspace{-2mm}
 \parbox{\halftext}{
\epsfxsize=7.5cm
\epsfbox{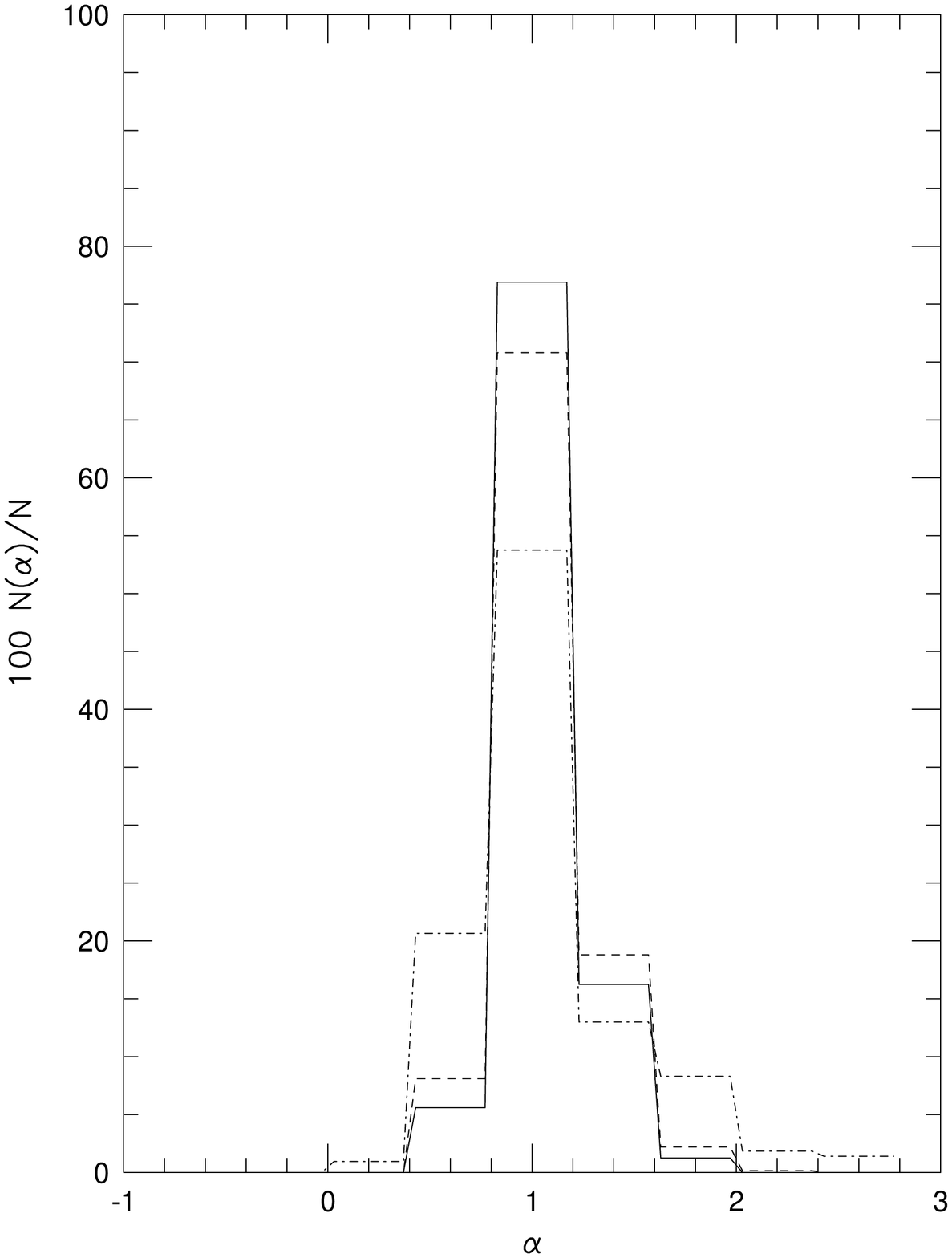}
   \caption{The percentage ($100 N(\alpha)/ N)$ of the distribution 
of $\alpha$ in bins with the interval $\Delta \alpha = 0.4$, for 
$D_{\rm lA}$ in the lens model 2 and model O with (0.2, 0).
  The lines have the same meaning as in Fig. 1. }}
\end{figure}
   
\begin{figure}[htb]
 \parbox{\halftext}{
\epsfxsize=7.5cm
\epsfbox{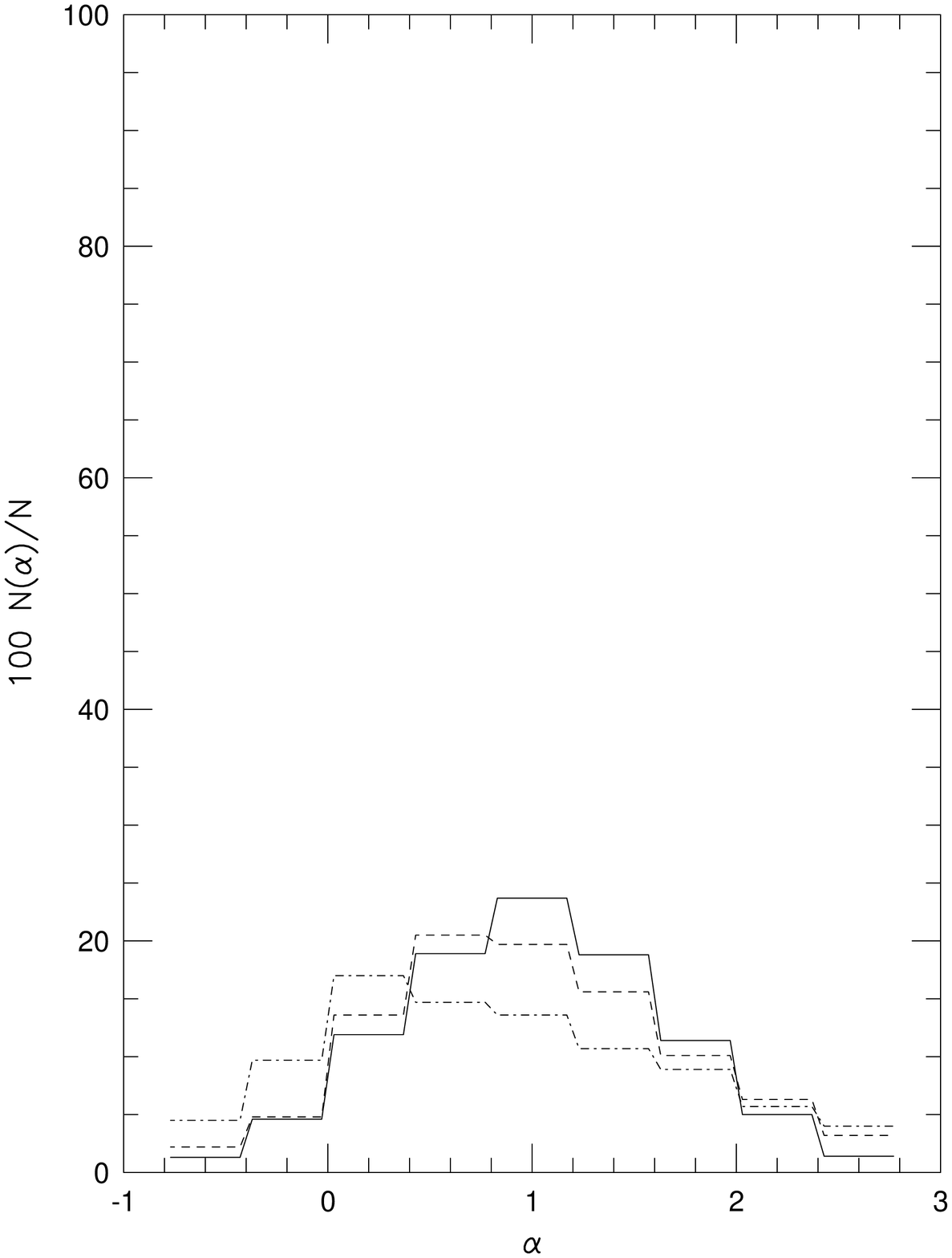}
   \caption{The percentage ($100 N(\alpha)/ N)$ of the distribution 
of $\alpha$ in bins with the interval $\Delta \alpha = 0.4$, for 
$D_{\rm aA}$ in the lens model 1 and model O with $(0.2, 0)$. 
The lines have the same meaning as in Fig. 1. }}
\hspace{-2mm}
 \parbox{\halftext}{
\epsfxsize=7.5cm
\epsfbox{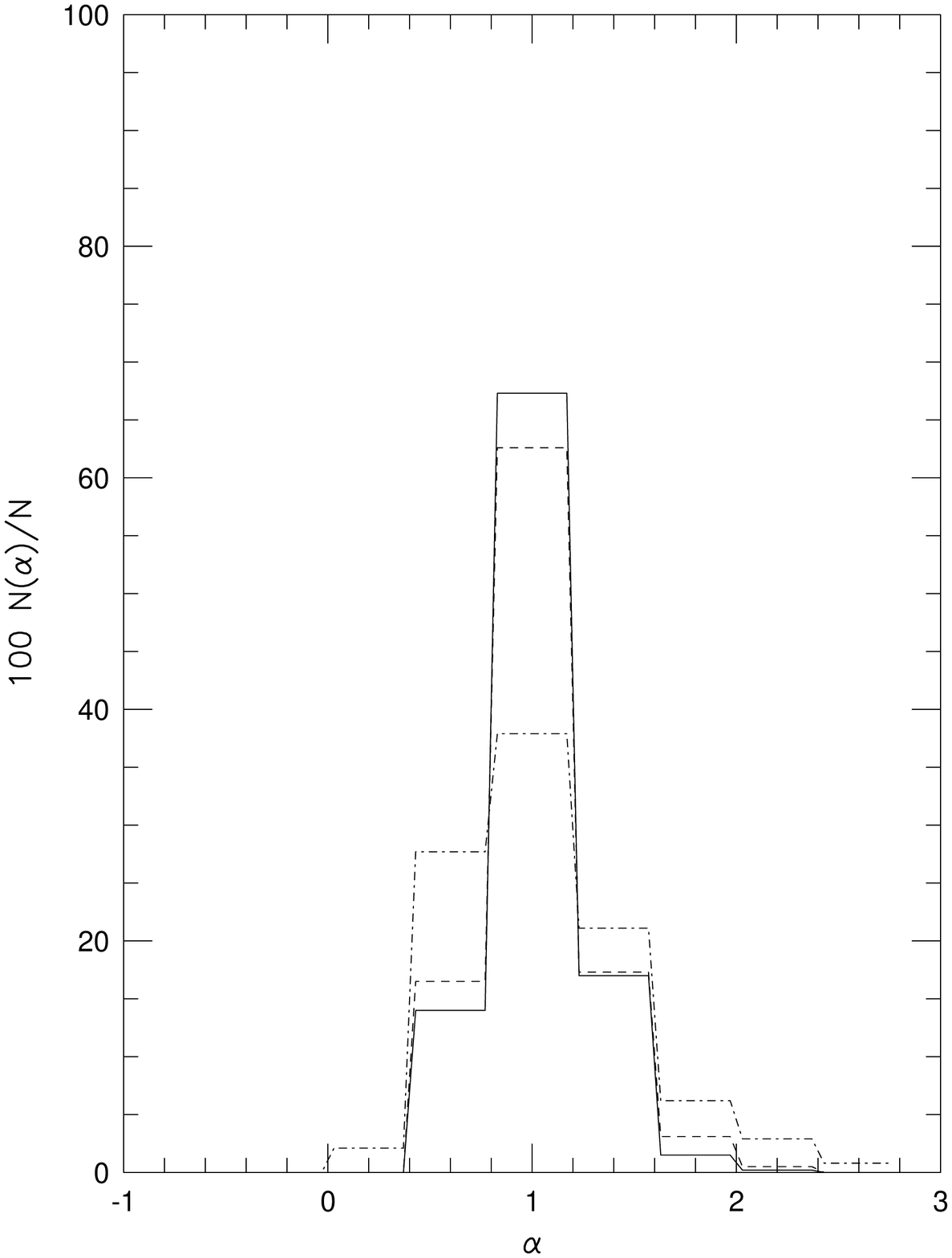}
   \caption{The percentage ($100 N(\alpha)/ N)$ of the distribution 
of $\alpha$ in bins with the interval $\Delta \alpha = 0.4$, for 
$D_{\rm aA}$ in the lens model 2 and model L with (0.2, 0).
  The lines have the same meaning as in Fig. 1. }}
\end{figure}

\begin{table}
\caption{The average clumpiness parameter $\bar{\alpha}$ and its dispersion 
$\sigma_\alpha$, and the average normalized distance $\bar{d}_{\rm A}$
and its dispersion $\sigma_d$ for two lens models
 in model O with $(\Omega_0, \lambda_0) = (0.2, 0)$. \
}

\label{table:3}
\begin{center}
\begin{tabular}{cccccccccc} \hline \hline
& & &linear & $(D_{\rm lA})$ & & &area & $(D_{\rm aA})$ & \\ \hline
lens & $z$ &$\bar{\alpha} $ & $\sigma_\alpha$ & $\bar{d}_{\rm A}
 $ & $\sigma_d$&$\bar{\alpha} $ & $\sigma_\alpha$ & $\bar{d}_{\rm A}
 $ & $\sigma_d$
\\ \hline
 &0.5& $1.02$ &  $1.99$ &  1.00 &  0.020 &$1.07$ &  $1.24$ &  1.00 &  0.012\\
 &1&   $1.07$ &  $1.55$ &  1.00 &  0.046 &$1.12$ &  $1.22$ &  1.00 &  0.037\\
1&2&   $1.06$ &  $1.06$ &  1.00 &  0.084 &$1.09$ &  $0.84$ &  0.99 &  0.067\\
 &3&   $1.04$ &  $0.89$ &  1.00 &  0.113 &$1.09$ &  $0.74$ &  0.99 &  0.094\\
 &4&   $1.03$ &  $0.82$ &  0.99 &  0.137 &$1.08$ &  $0.68$ &  0.99 &  0.115\\
 &5&   $1.03$ &  $0.75$ &  0.99 &  0.156 &$1.07$ &  $0.62$ &  0.99 &  0.129\\
\hline
 &0.5& $1.08$ &  $0.61$ &  1.00 &  0.006 &$1.04$ &  $0.44$ &  1.00 &  0.004\\
 &1&   $1.08$ &  $0.49$ &  1.00 &  0.014 &$1.08$ &  $0.41$ &  1.00 &  0.012\\
2&2&   $1.04$ &  $0.31$ &  1.00 &  0.025 &$1.06$ &  $0.26$ &  1.00 &  0.021\\
 &3&   $1.03$ &  $0.26$ &  1.00 &  0.033 &$1.06$ &  $0.22$ &  0.99 &  0.028\\
 &4&   $1.02$ &  $0.24$ &  1.00 &  0.040 &$1.05$ &  $0.20$ &  0.99 &  0.034\\
 &5&   $1.02$ &  $0.22$ &  1.00 &  0.046 &$1.05$ &  $0.18$ &  0.99 &  0.038\\
\hline
\end{tabular}
\end{center}
\end{table}
\bigskip

As a result of statistical analysis for this ray-shooting, we
derived the average clumpiness parameter $\bar{\alpha}$, the average 
normalized distance $\bar{d}_{\rm A}$, their dispersions 
($\sigma_\alpha$ and $\sigma_d$) and the distribution ($N(\alpha)$) of
$\alpha$. Here, in order to study the frequency of ray pairs with
$\alpha$, we consider many bins with the interval $\Delta \alpha =
0.4$ and the centers $\alpha_i = 1.0 \pm 0.4 i \ (i = 0, 1, 2, ...)$.
The number of ray pairs with $\alpha_i - \Delta \alpha/2 \leq \alpha 
\leq \alpha_i + \Delta \alpha/2$ is expressed as $N (\alpha_i)$ and the total
number of ray pairs is $N (= \sum_i N(\alpha_i))$.  
In Tables I $\sim$ III, $(\bar{\alpha}, \sigma_\alpha,
\bar{d_{\rm A}}, \sigma_d)$ for $D_{\rm lA}$ and $D_{\rm aA}$
in the two lens models are shown for models S, L and O, respectively. 
In Figs. 1 $\sim$ 12, the percentages of the distribution of $\alpha$, 
that is, $100 N(\alpha)/ N$ for $D_{\rm lA}$ and $D_{\rm aA}$ 
are shown for the above three models. 
The bar graphs in these figures have the same meaning as the line
graphs which were used in the previous paper.\cite{rf:tom98c}
In Figs. 1, 5 and 9, the distributions for six values of $z$ are
shown, and in the other figures those for only $z = 1, 3$ and $5$ are 
shown to avoid confusion.
The following types of statistical behavior are found from these 
tables and figures :

\medskip

\noindent (1) In all cases, both the average values $\bar{\alpha}$ and 
$\bar{d}_{\rm A}$ 
are nearly equal to $1$, so that the average angular distance can be
regarded as the Friedmann distance.

\noindent (2) For each angular distance, the two kinds of 
dispersions have different
behavior: $\sigma_\alpha$ increases in the order of O, L and S for the
same value of $z$, and $\sigma_\alpha$ in a given universe model 
decreases with the 
increase of $z$. On the other hand, $\sigma_d$ increases in the order
of L, O and S, and $\sigma_\alpha$ in a given universe model increases with the 
increase of $z$. This behavior is connected with the situation that 
the change in the Dyer-Roeder distance corresponding to the change in
$\alpha$ is small for $\alpha < 1$. 

\noindent (3) Generally the dispersions for $D_{\rm lA}$ are larger
than those for $D_{\rm aA}$, and the ratios of two dispersions are
$\sim 1.2$ for model S and $\sim 1.6$ for models L and O.
These differences can be seen also by comparing Figs. 1 and 3, Figs. 2 and
4, ..., and Figs. 10 and 12. 

\noindent (4) The dispersions in the lens model 2 are smaller than
those in the lens model 1. The ratios of the former to the latter are
$ \sim 2/3, 1/3, 1/3$ for universe models S, L and O, respectively.
 These differences can be seen similarly by comparing Figs. 1 and 2, 
Figs. 3 and 4, ..., and Figs. 11 and 12.
  In the two lens models of model S and in the lens model 2 of models L
and O, the angular diameter distances can be regarded as the Friedmann 
distance, because of the small dispersions. In the lens model 1 of
models L and O, however, we cannot always use the Friedmann distance,
because of comparatively large dispersions.

Using numerical ray-shooting in the $N$-body-simulating clumpy
cosmological models, we studied the statistical behavior of the
angular diameter distances $D_{\rm lA}$ and $D_{\rm aA}$ and
determined the clumpiness parameter
$\alpha$ by comparing it with the Friedmann distance $(\alpha = 1)$ and 
the Dyer-Roeder distance $(\alpha > 0)$. Moreover, we studied the
behavior of the normalized distance $d_{\rm A}$. The results show that 
all average values of $\alpha$ are nearly equal to 1, the
dispersions of the linear distance are slightly larger than those of
the area distance, and in the lens model 1 of
models L and O, the dispersions are not so small, and we cannot use
the Friedmann distance, while in the other cases we can use the
Friedmann distance because of small dispersions. 

In the above averaging process, all light rays were taken into
account. If we consider only weakly deflected light rays as
contributing to weak lensing, the dispersion  $\sigma_\alpha$ will 
be  slightly smaller than the values in the above tables. However, the
contribution of strong lensing to $\sigma_\alpha$ is small because of
its small frequency. 

Finally, we touch on the estimate of lensing correction to source
magnitudes based on Tables I $\sim$ III. Using $\sigma_d$ in the area
distance it is given by
\begin{equation}
  \label{eq:d8}
\Delta m (z) = {5 \over 2} \log [1 + 2 \sigma_d] = 2.18 \sigma_d
\end{equation}
for the source at $z$. Its values for the lens model $(1, 2)$ are 

 $\Delta m (0.5) = (0.033, 
0.020), (0.026, 0.009)$ and $(0.026, 0.009),$ 

 $\Delta m (1.0) = 
(0.061, 0.037), (0.074, 0.022)$ and $(0.081, 0.026)$

\noindent for models S, L and O, respectively. This $\Delta m$ (for
the separation angle $\theta = 1$ arcsec) was derived independently of 
$\Delta m$ (for the separation angle $\theta = 2$ arcsec), shown in the
\S 3 of Tomita, Premadi and Nakamura's paper in this Supplement, 
but their values at $z = 1$ are roughly consistent.
The lensing correction to source magnitudes was also investigated by
Holz in different lens models and inhomogeneous models.\cite{rf:holz98}

\bigskip
\section{Shear effect on distances}
As we have shown in \S 2, the evolution of a cross sectional
area of a light ray bundle is determined by Ricci and Weyl focusing
along the trajectory of the ray bundle.
Ricci focusing is a convergency effect due to matter in the ray
bundle. On the other hand, Weyl focusing is a result of the tidal
shear on the ray bundle induced by the inhomogeneous distribution of
matter.

One of the main difficulties in deriving a distance-redshift relation
analytically for a realistic inhomogeneous universe lies in estimating 
the effect of Weyl focusing  on ray bundles.

Since the pioneering work of Gunn, \cite{gu67} there has been a great
deal of progress in constructing
an analytical approach to investigate statistical quantities of
realistic distances
(e.g., dispersion and skewness of probability distribution of image 
magnifications). 
Babul and Lee, \cite{ba91} among others, examined lensing magnification
effects on distances due to the large scale structures
($\geq 0.5h^{-1}$Mpc, where Hubble constant $H_0 = 100 h$km/sec/Mpc).
They found that the dispersion in image magnifications is negligible
even for sources at a redshift of 4.
At the same time, they pointed out that the dispersion is
very sensitive to the nature of the matter distribution on small
scales.
In their study, the effects of Weyl focusing were neglected based,
on a numerical study by Jaroszy\'nski et al.\ \cite{ja90} in which the
lensing magnification effects due to large scale structure ($\geq
1h^{-1}$Mpc) in cold dark matter models were examined, 
and it was concluded that Weyl focusing has no significant effect on
image magnifications.
Frieman \cite{fr97} improved Babul and Lee's study to reflect the recent
developments in numerical and observational studies of large scale
structure.
Although he took small scale nonlinear structure into account,
Weyl focusing effects were neglected without any reasonable
basis.

Nakamura \cite{rf:naka97}  examined the effects of shear on image
magnification
in the cold dark matter model universe with a linear density perturbation.
He found that the effect is sufficiently small and concluded that
Weyl focusing can be safely neglected for a light ray passing through
linear density inhomogeneities. 

The above cited studies mainly focus on large scale inhomogeneities, 
whereas the effects of small scale objects, such as galaxies and
clusters of galaxies, have not been satisfactorily taken into account.
It is, however, not clear whether the Weyl focusing effect due
to small scale inhomogeneities has a significant effect on
distances.
In this section, we discuss the Weyl focusing effect due
to small scale inhomogeneities, mainly following Hamana. \cite{ha99}

\subsection{Basic equations}
We first derive an evolution equation of lensing magnification from
the null geodesic equation, \cite{ha98} which is equivalent to the
optical scalar
equations,\cite{rf:sasaki93,se94} and is convenient to examine 
gravitational lensing  effects. 

We rewrite the cosmological Newtonian metric (\ref{eq:int2}) as
\begin{equation}
\label{conf-metric}
ds^2 = a^2(\eta)\left[ -(1+2 \Phi) d\eta^2 + (1-2 \Phi) \gamma_{ij}
dx^i dx^j \right],
\end{equation}
where $\eta$ is a conformal time: $d\eta \equiv cdt$.
We write the above metric as $g_{\mu \nu} = a^2 \hat{g}_{\mu \nu}$. 
Since the light cone structure is invariant
under the conformal transformation of the metric, in the following we
work in conformally related $\hat{g}$ world.

Let us consider an infinitesimal bundle of light rays
intersecting at the observer. We denote a connecting vector
which connects the fiducial light ray $\gamma$ to one of its neighbors 
as $\xi^\mu$.
All gravitational focusing and shearing effects on the infinitesimal
light ray bundle are described by the geodesic deviation equation,
\begin{equation}
\label{geodev}
{{d^2 \xi^{\mu}} \over {d \lambda^2}} = -{R^{\mu}}_{\alpha \nu \beta}
\xi^{\nu} k^{\alpha} k^{\beta},
\end{equation}
where $k^{\alpha}=dx^{\alpha}/d\lambda$, and $\lambda$ is the affine
parameter along the fiducial light ray $\gamma$. We introduce a dyad
basis $e_{\mu}{}^A$ ($A$, $B$, $C,...=1,2$) in the two-dimensional screen
orthogonal to $k^{\mu}$ and parallel-propagated along $\gamma$.
The screen components of the connection vector are given by
\begin{equation}
\label{Y}
Y^A = e_{\mu}{}^A \xi^{\mu}.
\end{equation} 
{}From the geodesic deviation equation (\ref{geodev}), one can
immediately find that $Y^A$ satisfies the Jacobi differential equation
\begin{equation}
\label{jaco}
{{d^2 Y_A}\over {d\lambda^2}} = {\cal{T}}_{A B} Y^B,
\end{equation}
where ${\cal{T}}_{AB}$ is the so-called optically tidal
matrix.~\cite{se94} From the metric (\ref{conf-metric}), up to first
order in $\Phi$, this matrix is given by 
\begin{equation}
\label{tid}
\mbox{{\boldmath${\cal{T}}$}} = 
-K \mbox{{\boldmath${\cal{I}}$}} - \left(
\begin{array}{ll}
{\cal{R}} + \mbox{Re}\left[{\cal{F}}\right] &
\mbox{Im}\left[{\cal{F}}\right] \\
\mbox{Im}\left[{\cal{F}}\right] &
{\cal{R}} - \mbox{Re}\left[{\cal{F}}\right]
\end{array}
\right),
\end{equation}
where {\boldmath${\cal{I}}$} is the $2 \times 2$ identity matrix, and 
${\cal{R}}$ and ${\cal{F}}$ represent the Ricci and Weyl focusing
induced by the density inhomogeneities, respectively:
\begin{eqnarray}
\label{Ricci}
{\cal{R}} &=&  \Delta^{(3)} \Phi ={{3 \Omega_0 H_0^2}
\over {2}} {{\delta} \over a},\\
\label{Wely}
{\cal{F}} &=&  \Phi_{,11}-\Phi_{,22} + 2 i
\Phi_{,12} . 
\end{eqnarray}
Here $\Delta^{(3)}$ is the Laplacian operator in the spatial section,
and $\delta$ is the density contrast defined by $\delta \equiv
(\rho-\rho_b)/\rho_b$, where $\rho_b$ is the mean matter density.
Owing to the linearity of (\ref{jaco}), 
the solution of $Y^A$ can be written in terms of its initial value
$dY^A/d\lambda |_{\lambda=0} = \vartheta^A$ and the $\lambda$-dependent
linear transformation matrix ${\cal{D}}_{AB}$ can be written as 
\begin{equation}
\label{D}
Y^A(\lambda) = {\cal{D}}^A{}_B(\lambda) \vartheta^B.
\end{equation}
Substituting the last equation into the Jacobi differential equation
(\ref{jaco}), we obtain 
\begin{equation}
\label{evoD}
{{d^2{\cal{D}}_{AB}} \over {d \lambda^2}} = {\cal{T}}_{AC} {\cal{D}}_{CB}.
\end{equation}

Now, we derive an evolution equation of the lensing magnification
matrix relative to the smooth Friedmann distance from (\ref{evoD}). 
First, we write (\ref{tid}) as
{\boldmath${\cal{T}}$}$ = ${\boldmath${\cal{T}}$}${}^{(0)} +
${\boldmath$\delta{\cal{T}}$}, with {\boldmath${\cal{T}}$}${}^{(0)} =
-K${\boldmath${\cal{I}}$}, and {\boldmath$\delta {\cal{T}}$} is the
second term in (\ref{tid}). In the homogeneous case,
{\boldmath$\delta{\cal{T}}$} is vanishing, and the solution of
{\boldmath${\cal{D}}$} is ${\cal{D}}_{AB}(\lambda) = D_f(\lambda)
\delta_{AB}$, where $D_f$ is, of course, the
standard angular diameter distance in the background Friedmann
universe. 

It is natural to define the lensing magnification matrix relative to the
corresponding Friedmann universe as
\begin{equation}
\label{defM}
{\cal{M}}_{AB} (\lambda) \equiv {{{\cal{D}}_{AB} (\lambda)} \over {D_f
(\lambda)}} .
\end{equation}
Differentiating ${\cal{M}}_{AB}$ twice with respect to $\lambda$ and
using (\ref{evoD}), one finds
\begin{equation}
\label{difM}
{{d^2{\cal{M}}_{AB}} \over {d\lambda^2}} = -{2 \over {D_f}} {{d D_f}
\over {d \lambda}} {{d {\cal{M}}_{AB}} \over {d\lambda}} + \delta
{\cal{T}}_{AC} {\cal{M}}_{CB}.
\end{equation}
With the initial conditions {\boldmath${\cal{M}}$}$(\lambda) |_{\lambda=0} =
${\boldmath${\cal{I}}$} and $d${\boldmath${\cal{M}}$}
$(\lambda)/{d\lambda} |_{\lambda=0} = ${\boldmath${\cal{O}}$}, \cite{se94} 
the last equation can be written in the integral form 
\begin{equation}
\label{Mab}
{\cal{M}}_{AB} (\lambda) = \delta_{AB} + \int_0^{\lambda} d \lambda'
{{D_f(\lambda-\lambda') D_f(\lambda')} \over {D_f(\lambda)}} \delta
{\cal{T}}_{AC}(\lambda') {\cal{M}}_{CB} (\lambda'). 
\end{equation}
This is the general form of the evolution equation of the lensing magnification
matrix relative to the Friedmann distance in multiple gravitational
lensing theory.~\cite{rf:sef}
Note, in general, this equation is not an explicit equation for
${\cal{M}}_{AB}$, since it involves an integration over the
optical tidal matrix evaluated 
on the light ray path, such that one first has to solve a null
geodesic equation. 
Since, for almost all cases of cosmological interest, the
deflection angle is very small,\cite{rf:fs89}  
we will neglect the deflection of
light rays.

\subsection{Order-of-magnitude estimate}
We now examine the magnitude of lensing effects on light ray bundles due 
to randomly distributed virialized objects (e.g.\ galaxies and
clusters of galaxies) adopting an order-of-magnitude
estimate. \cite{rf:fs89,pe93} 
As can be seen in Eq. (\ref{Mab}), if the magnitude of the
components of the matrix,
$\int d \lambda ' D_f(\lambda-\lambda') D_f(\lambda')/ D_f(\lambda) \delta
${\boldmath$\cal{T}$}$(\lambda')$ is small, 
the magnitude of the lensing effects is dominated by these terms.
We, therefore, examine the magnitude of these terms.
For simplicity, we denote these terms as $\delta {\cal{M}}$. 

Supposing that lensing objects are randomly distributed and that each
has a mass  $M = 2 {\sigma_v}^2 l /G$, 
where ${\sigma_v}$ is the one-dimensional velocity dispersion of the
lens objects, and $l$ is a characteristic comoving size of a lens
object.
Hence the mean comoving number density of the lens objects is 
\begin{equation}
n_L ={{3 \Omega_L {H_0}^2} \over {16 \pi}} {1\over{\sigma_v}^{2}}
{1\over {l}},
\end{equation}
where
$\Omega_L$ is the density parameter of lens objects defined by
$\Omega_L \equiv \rho_L/(3 H_0^2/8 \pi G)$, with $\rho_L$ is the mean
density of the lens objects.
Thus, the mean comoving separation distance is
\begin{equation}
r_0 = \left({{ 16 \pi} \over {3 \Omega_L H_0^2}}  \sigma_v^2 l
\right)^{1\over 3}.
\end{equation}
Then for a geodesic affine comoving distance of $\lambda$, the light
ray gravitationally encounters such objects
$N_g = \lambda/r_0$
times on average.
At each encounter, the contribution to the lensing magnification matrix is 
\begin{eqnarray}
\Delta {\cal{M}} &=& 4\pi \left( {{\sigma_v} \over c} \right)^2
{{r_0} \over {b^2}} {{D_f(\lambda_d)
D_f(\lambda_s-\lambda_d)} \over {D_f(\lambda_s)}}\nonumber\\
&\sim& 4\pi \left( {{\sigma_v} \over c} \right)^2
{1 \over {r_0}} {{D_f(\lambda_d)
D_f(\lambda_s-\lambda_d)} \over {D_f(\lambda_s)}},
\end{eqnarray}
where $D_f(\lambda_i-\lambda_j)$ is the comoving angular diameter
distance, the subscripts $d$ and $s$ indicate the lens and source,
respectively, and $b$ is the comoving impact parameter.
In the above expression, we have assumed that the mean comoving impact
parameter is of order $r_0$.
Since the sign of each contribution will be random, the total
contribution to the lensing magnification matrix is
\begin{eqnarray}
\delta {\cal{M}} &\sim&
\Delta {\cal{M}} \sqrt{N_g}\nonumber\\ 
&\sim& 
\sqrt{3 \over 4} \sqrt{\Omega_L} \left[ 4\pi
\left ({{\sigma_v} \over c}\right)^2 {1\over {l}} {c \over {H_0}}
\right]^{1\over 2} \left\langle{{{H_0} \over c} {{{D_f(\lambda_d)
D_f(\lambda_s- \lambda_d)}}
\over {D_f(\lambda_s)}}
}\right\rangle
\left[{{H_0} \over c} \lambda \right]^{1\over 2}\nonumber\\
\label{magoft}
&\sim& \sqrt{\Omega_L} \sqrt{\nu} 
\left\langle{{{H_0} \over c} {{{D_f(\lambda_d)
D_f(\lambda_s- \lambda_d)}}
\over {D_f(\lambda_s)}}
}\right\rangle
\left[{{H_0} \over c} \lambda \right]^{1\over 2},
\end{eqnarray}
where $\nu$ is a compactness parameter of a lens object defined by
$\nu \equiv 4 \pi (\sigma_v/c)^2 l^{-1} \\
\times (c/H_0)$.
The contribution from direct encounters can be similarly
estimated by noting that the average number of encounters is
\begin{equation}
N_d ={{l^2 \lambda} \over {r_0^3}} = {{3 \Omega_L {H_0}^2}
\over {16 \pi}} {1\over{\sigma_v}^2} l \lambda,
\end{equation}
with each encounter contributing
\begin{equation}
\Delta {\cal{M}}_d = 4\pi \left({{\sigma_v} \over c} \right)^2
{1\over l} {{D_f(\lambda_d) D_f(\lambda_s-\lambda_s)} \over
{D_f{\lambda_s}}}, 
\end{equation}
with random sign.
The result turns out to be the same as that of gravitational distant
encounters, given by Eq. (\ref{magoft}).  
The comoving affine distance $\lambda$ becomes $c/H_0$ at the source redshift 
$z_s \sim 3$, and the averaged value of the distance combination over the
lens redshifts is of order 
\begin{equation}
\left\langle {{H_0} \over c}
{{D_f(\lambda_d)D_f(\lambda_s-\lambda_d)}\over {D_f(\lambda_s)}}
\right\rangle \sim {\cal{O}}(0.1).
\end{equation}
Accordingly, we find that the magnitude of the total contribution of
the lensing effects to the lensing magnification matrix
scales as $\sim 0.1 \sqrt{\Omega_L} \sqrt{\nu}$ for the
source redshift $z_s \sim 3$. We have the relation 
$\Omega_L \leq \Omega_0$ by definition, and $\Omega_0$ appears to
be less than unity. Thus $\Omega_L \leq 1$.
On the other hand, $\nu \le 1$ for
galaxies and clusters of galaxies. 
Therefore a typical value of  gravitational lensing effects on the
lensing matrix can be expected to be $ {\cal{O}}(0.1)$ or
smaller for a majority of random lines of sight.

The lensing magnification factor of a point like image is defined by
the determinant of the lensing magnification matrix.
Taking the determinant of the lensing magnification matrix
(\ref{Mab}), and expanding it in powers of  $\delta{\cal{M}}$,
one can easily find that the leading term of Weyl focusing effects
is of order $\delta{\cal{M}}^2$.
On the other hand, that of the Ricci focusing term is of order
$\delta{\cal{M}}$. 
Since we have seen that a typical value of $\delta {\cal{M}}$ is expected to 
be ${\cal{O}}(0.1)$ or smaller, we
can conclude that, at least from a statistical point of view,  Weyl
focusing has no significant effects on the image
magnifications or equivalently on the distances.

\subsection{Numerical investigation}
The above argument may sound too naive.
One of the authors (T.H.), numerically investigated Weyl focusing
effects on image magnifications by using the multiple
gravitational lens theory.\cite{ha99}
He focused on gravitational lensing effects due to small scale
virialized objects, such as galaxies and clusters of galaxies.
He considered a simple model of an inhomogeneous universe.
The matter distribution in the universe was modeled by randomly
distributed isothermal objects.
He found that, for the majority of the random lines of sight, Weyl
focusing has no significant effect, and the image magnification of
a point like source within a redshift of 5 is dominated by Ricci
focusing. 
He also found that his result agrees well with the 
order-of-magnitude estimate given above.

To summarize, we conclude that, except for a statistically very rare
kind of light ray, Weyl focusing has no significant effect on 
image magnifications or equivalently on the distances.

\section{Concluding remarks}
Lensing observation in inhomogeneous universes was discussed 
in \S 3, based on the so-called Dyer-Roeder distance, in which 
one of the main assumptions is neglecting Weyl focusing. 
Such a neglection seems correct in our universe, as was shown in 
the \S 5. 

The average angular distances in inhomogeneous model universes are the 
Friedmann distances, as was shown in \S 4, but individual ray 
bundles have various values of clumpiness parameters $\alpha$ because
of their dispersions. The observational quantities are sensitively
dependent on $\alpha$, as was shown in the \S 3, and so they may 
have dispersions similar to $\alpha$.

The difference between linear and area angular diameter distances, which
is caused by the shear, is generally small in accord with the result 
in \S 5, even though we considered small-scale
inhomogeneities, but the difference in the low-density models is
found to be larger than that in the Einstein-de Sitter model. This 
implies that the shear effect is comparatively larger in the low-density 
models.

\section*{Acknowledgements}
K.T. would like to thank Y.~Suto for helpful discussions about
$N$-body simulations. His numerical computations were performed on 
the YITP computer system.
H.A.\ and T.H.\ would like to thank T.~Futamase for fruitful discussions.
H.A. would like to thank M.~Kasai, M.~Sasaki and T.~Tanaka 
for useful conversations.

\end{document}